\documentclass[
  journal=pasa,
  manuscript=research-paper, %% or "review"
  year=2025,
  volume=,
]{cup-journal}

\usepackage{microtype,siunitx,booktabs,physics, multicol,orcidlink}

\usepackage{subcaption}
\usepackage{amsfonts, amssymb, amsmath}

\makeatletter
\DeclareRobustCommand{\HI}{%
  \mbox{H\check@mathfonts\fontsize\sf@size\z@\selectfont I}%
}
\DeclareRobustCommand{\HII}{%
  \mbox{H\check@mathfonts\fontsize\sf@size\z@\selectfont II}%
}
\makeatother

\DeclareMathOperator{\argmax}{argmax}

\title{The MAGPI Survey: the kinematic morphology-density relation (or lack thereof) and the Hubble sequence at $z\sim0.3$}

\author{Caroline Foster \orcidlink{https://orcid.org/0000-0003-0247-1204}}
    \affiliation{School of Physics, University of New South Wales, Sydney, NSW 2052, Australia}
    \alsoaffiliation{ARC Centre of Excellence for All Sky Astrophysics in 3 Dimensions (ASTRO 3D)}
    \email[Caroline Foster]{c.foster@unsw.edu.au}
\author{Mark W. Donoghoe}
    \affiliation{Clinical Research Unit, Medicine and Health, UNSW Sydney, Sydney, NSW, Australia}
    \alsoaffiliation{Kirby Institute, Medicine and Health, UNSW Sydney, Sydney, NSW, Australia}
    \alsoaffiliation{Stats Central, Mark Wainwright Analytical Centre, UNSW Sydney, Sydney, NSW, Australia}
\author{Andrew Battisti \orcidlink{https://orcid.org/0000-0003-4569-2285}}
    \affiliation{International Centre for Radio Astronomy Research, The University of Western Australia, 35 Stirling Highway, Crawley WA 6009, Australia}
    \affiliation{Research School of Astronomy and Astrophysics, Australian National University, Cotter Road, Weston Creek, ACT, 2611, Australia}
    \alsoaffiliation{ARC Centre of Excellence for All Sky Astrophysics in 3 Dimensions (ASTRO 3D)}
\author{Francesco D'Eugenio \orcidlink{https://orcid.org/0000-0003-2388-8172}}
    \affiliation{Kavli Institute for Cosmology, University of Cambridge, Madingley Road, Cambridge, CB3 0HA, United Kingdom}
    \alsoaffiliation{Cavendish Laboratory - Astrophysics Group, University of Cambridge, 19 JJ Thomson Avenue, Cambridge, CB3 0HE, United Kingdom}
\author{Katherine Harborne \orcidlink{https://orcid.org/0000-0002-2043-7985}}
    \affiliation{International Centre for Radio Astronomy Research, The University of Western Australia, 35 Stirling Highway, Crawley WA 6009, Australia}
    \alsoaffiliation{ARC Centre of Excellence for All Sky Astrophysics in 3 Dimensions (ASTRO 3D)}
\author{Thomas Venville\, \orcidlink{0000-0003-0278-9933}}
    \affiliation{Research School of Astronomy and Astrophysics, Australian National University, Cotter Road, Weston Creek, ACT, 2611, Australia}
\author{Claudia Del P. Lagos \orcidlink{https://orcid.org/0000-0003-3021-8564}}
    \affiliation{International Centre for Radio Astronomy Research, The University of Western Australia, 35 Stirling Highway, Crawley WA 6009, Australia}
    \alsoaffiliation{ARC Centre of Excellence for All Sky Astrophysics in 3 Dimensions (ASTRO 3D)}
\author{J. Trevor Mendel \orcidlink{https://orcid.org/0000-0002-6327-9147}}
    \affiliation{Research School of Astronomy and Astrophysics, Australian National University, Cotter Road, Weston Creek, ACT, 2611, Australia}
    \alsoaffiliation{ARC Centre of Excellence for All Sky Astrophysics in 3 Dimensions (ASTRO 3D)}
\author{Ryan Bagge \orcidlink{https://orcid.org/0009-0002-2753-3248}}
    \affiliation{School of Physics, University of New South Wales, Sydney, NSW 2052, Australia}
    \alsoaffiliation{ARC Centre of Excellence for All Sky Astrophysics in 3 Dimensions (ASTRO 3D)}
\author{Stefania Barsanti \orcidlink{https://orcid.org/0000-0002-9332-5386}}
    \affiliation{Research School of Astronomy and Astrophysics, Australian National University, Cotter Road, Weston Creek, ACT, 2611, Australia}
    \alsoaffiliation{ARC Centre of Excellence for All Sky Astrophysics in 3 Dimensions (ASTRO 3D)}
\author{Sabine Bellstedt \orcidlink{https://orcid.org/0000-0003-4169-9738}}
    \affiliation{International Centre for Radio Astronomy Research, The University of Western Australia, 35 Stirling Highway, Crawley WA 6009, Australia}
\author{Alina Boecker \orcidlink{https://orcid.org/0000-0003-2945-106X}}
    \affiliation{Instituto de Astrofísica de Canarias, C/ Vía Láctea s/n, E-38205 La Laguna, Spain}
    \alsoaffiliation{Departamento de Astrofisica, Universidad de La Laguna, E-38200 La Laguna, Tenerife, Spain}
    \alsoaffiliation{Department of Astrophysics, University of Vienna, Türkenschanzstraße 17, 1180, Vienna, Austria}
\author{Qianhui Chen}
    \affiliation{Research School of Astronomy and Astrophysics, Australian National University, Cotter Road, Weston Creek, ACT, 2611, Australia}
    \alsoaffiliation{ARC Centre of Excellence for All Sky Astrophysics in 3 Dimensions (ASTRO 3D)}
\author{Caro Derkenne \orcidlink{https://orcid.org/0000-0003-3474-3542}}
    \affiliation{School of Mathematical and Physical Sciences, Macquarie University, NSW 2109, Australia}
    \alsoaffiliation{ARC Centre of Excellence for All Sky Astrophysics in 3 Dimensions (ASTRO 3D)}
\author{Anna Ferr\'e-Matteu \orcidlink{https://orcid.org/0000-0002-6411-220X}}
    \affiliation{Instituto de Astrofísica de Canarias, C/ Vía Láctea s/n, E-38205 La Laguna, Spain}
    \alsoaffiliation{Departamento de Astrofisica, Universidad de La Laguna, E-38200 La Laguna, Tenerife, Spain}
\author{Eda Gjergo \orcidlink{https://orcid.org/0000-0002-7440-1080}}
    \affiliation{School of Astronomy and Space Science, Nanjing University, Nanjing 210093, People's Republic of China}
    \alsoaffiliation{Key Laboratory of Modern Astronomy and Astrophysics ( Nanjing University ), Ministry of Education, Nanjing 210093, People's Republic of China}
\author{Anshu Gupta \orcidlink{https://orcid.org/0000-0002-8984-3666}}
    \affiliation{International Centre for Radio Astronomy Research (ICRAR), Curtin University, Bentley, WA, Australia}
    \alsoaffiliation{ARC Centre of Excellence for All Sky Astrophysics in 3 Dimensions (ASTRO 3D)}
\author{Eric G. M. Muller \orcidlink{https://orcid.org/0000-0001-5621-1577}}
    \affiliation{Research School of Astronomy and Astrophysics, Australian National University, Cotter Road, Weston Creek, ACT, 2611, Australia}
    \alsoaffiliation{ARC Centre of Excellence for All Sky Astrophysics in 3 Dimensions (ASTRO 3D)}
\author{Giulia Santucci \orcidlink{https://orcid.org/0000-0003-3283-4686}}
    \affiliation{International Centre for Radio Astronomy Research, The University of Western Australia, 35 Stirling Highway, Crawley WA 6009, Australia}
    \alsoaffiliation{ARC Centre of Excellence for All Sky Astrophysics in 3 Dimensions (ASTRO 3D)}
\author{Hye-Jin Park \orcidlink{https://orcid.org/0000-0002-9809-6631}}
    \affiliation{Research School of Astronomy and Astrophysics, Australian National University, Cotter Road, Weston Creek, ACT, 2611, Australia}
    \alsoaffiliation{ARC Centre of Excellence for All Sky Astrophysics in 3 Dimensions (ASTRO 3D)}
\author{Rhea-Silvia Remus \orcidlink{https://orcid.org/0009-0008-9260-7278}}
    \affiliation{Universitäts-Sternwarte München, Fakultät für Physik, Ludwig-Maximilians Universität, Scheinerstr. 1, D-81679 München, Germany}
\author{Sabine Thater \orcidlink{https://orcid.org/0000-0003-1820-2041}}
    \affiliation{Department of Astrophysics, University of Vienna, Türkenschanzstraße 17, 1180, Vienna, Austria}
\author{Jesse van de Sande \orcidlink{https://orcid.org/0000-0003-2552-0021}}
    \affiliation{School of Physics, University of New South Wales, Sydney, NSW 2052, Australia}
    \alsoaffiliation{ARC Centre of Excellence for All Sky Astrophysics in 3 Dimensions (ASTRO 3D)}
\author{Sam Vaughan \orcidlink{https://orcid.org/0000-0003-2265-7727}}
    \affiliation{School of Mathematical and Physical Sciences, Macquarie University, NSW 2109, Australia}
    \alsoaffiliation{ARC Centre of Excellence for All Sky Astrophysics in 3 Dimensions (ASTRO 3D)}
\author{Sarah Brough\,\orcidlink{0000-0002-9796-1363}}
    \affiliation{School of Physics, University of New South Wales, Sydney, NSW 2052, Australia}
    \alsoaffiliation{ARC Centre of Excellence for All Sky Astrophysics in 3 Dimensions (ASTRO 3D)}
\author{Scott M. Croom\,\orcidlink{0000-0003-2880-9197}}
    \affiliation{Sydney Institute for Astronomy, School of Physics, A28, The University of Sydney, NSW 2006, Australia}
    \alsoaffiliation{ARC Centre of Excellence for All Sky Astrophysics in 3 Dimensions (ASTRO 3D)}
\author{Lucas M. Valenzuela\,\orcidlink{0000-0002-7972-9675}}
    \affiliation{Universitäts-Sternwarte München, Fakultät für Physik, Ludwig-Maximilians Universität, Scheinerstr. 1, D-81679 München, Germany}
\author{Emily Wisnioski \orcidlink{https://orcid.org/0000-0003-1657-7878}}
    \affiliation{Research School of Astronomy and Astrophysics, Australian National University, Cotter Road, Weston Creek, ACT, 2611, Australia}
    \alsoaffiliation{ARC Centre of Excellence for All Sky Astrophysics in 3 Dimensions (ASTRO 3D)}
\author{the MAGPI Team}

\keywords{galaxies: kinematics and dynamics -- galaxies: evolution} %% First letter not capped

\usepackage{hyperref} 
\hypersetup{colorlinks,citecolor=blue,linkcolor=blue,urlcolor=blue}

\begin{document}

\begin{abstract}
This work presents visual morphological and dynamical classifications for 637 spatially resolved galaxies, most of which are at intermediate redshift ($z\sim0.3$), in the Middle-Ages Galaxy Properties with Integral field spectroscopy (MAGPI) Survey. For each galaxy, we obtain a minimum of 11 independent visual classifications by knowledgeable classifiers. We use an extension of the standard Dawid-Skene bayesian model introducing classifier-specific confidence parameters and galaxy-specific difficulty parameters to quantify classifier confidence and infer reliable statistical confidence estimates. Selecting sub-samples of 86 bright ($r<20$ mag) high-confidence ($>0.98$) morphological classifications at redshifts ($0.2 \le z \le0.4$), we confirm the full range of morphological types is represented in MAGPI as intended in the survey design. Similarly, with a sub-sample of 82 bright high-confidence stellar kinematic classifications, we find that the rotating and non-rotating galaxies seen at low redshift are already in place at intermediate redshifts. We \textit{do not} find evidence that the kinematic morphology-density relation seen at $z\sim0$ is established at $z\sim0.3$. We suggest that galaxies without obvious stellar rotation are dynamically pre-processed sometime before $z\sim0.3$ within lower mass groups before joining denser environments.
\end{abstract}

\section{Introduction}\label{sec:introduction}

Galaxies are routinely classified by visual inspection of their appearance. Astronomers have devised and improved classification schemes since the 1900's. Amongst the first and most popular classification schemes is that  devised by \citet[][the so-called ``tuning fork'']{Hubble26} and subsequent expansions suggested by \citet{deVaucouleurs59}, \citet{Sandage81} and others (see \citealt{Sandage05} for a historical review).

Visual classifications are not just helpful in grouping galaxies into families, but a broad body of literature has shown that these visual characteristics of galaxies encode important information about internal and external physical processes. This is evidenced by the fact that visual morphology has been linked to many other properties: colour \citep[e.g.][]{NedTaylor15, Correa17}, star formation rate \citep{Sandage86,Kennicutt98,Wuyts11}, galaxy interactions \citep{Lotz08,Kannan15}, environment \citep[][i.e. the morphology-density relation]{Dressler80,Goto03} and secular processes such as internal perturbations, torques from bars, spiral scattering, etc \citep[e.g.][]{Buta13, Sellwood14}.

The overall structural makeup of the galaxy population has been shown to have evolved both visually \citep[e.g.][]{Abraham01,Conselice14} and dynamically \citep[e.g.][]{Wisnioski15,Bezanson18,DEugenio23b}. This points to the importance of careful morphological and dynamical classifications of galaxies in studies comparing samples at different redshifts.

As such, the literature on visual classifications is broad and historically rich. Studies pertaining to visual classification of galaxies range from historical surveys of nearby galaxies \cite[e.g.][]{Nilson73,Sandage81,deVaucouleurs91} to modern large imaging surveys such as e.g. the Sloan Digital Sky Survey \citep[SDSS,][]{Abazajian09} and the Dark Energy Survey \citep{Cheng21a}. 

As samples of galaxies to visually classify have grown in proportions, the required individual efforts for obtaining reliable visual classification have become intractable. Astronomers are now turning to citizen science projects \citep[e.g.][]{Willett17,VazquezMata22,PorterTemple22} and routinely use automated classification techniques \citep[e.g.][]{HuertasCompany11,HuertasCompany15,DominguezSanchez19,Martin20,Cavanagh21,Cheng21b,Walmsley22a,Omori23,Desmons24} to handle the deluge of data.

%Sub-samples and more detailed analysis such as structure decomposition \citet{Khanday22}?

Morphological classification of galaxies at higher redshift is complicated by a number of confounding factors including the unavailability of similar rest-frame band filters, shallower relative depth and typically lower spatial resolution \citep[see e.g.][]{Ren24,Salvador24}. Visual classifications may thus need adapting for higher redshift samples \citep[e.g.][]{Masters11,Kartaltepe23,Tohill23, Conselice24}. 
Indeed, faint features such as tidal features \citet{Bilek20,Desmons23b}, as well as comparatively small features such as central bars or spiral arms \citep[e.g.][]{Masters21} may be more difficult to identify at high redshift depending on image quality and depth. High redshift galaxies (beyond $z\sim2$) may be morphologically and structurally different from their local counterparts, often not having bulges and discs \citep[e.g.][]{Sweet20}.

As with morphology, visual classification of the dynamics of galaxies through careful inspection of kinematic maps may help identify internal sub-structure, signatures of past interactions or secular processes and environmental effects.
As such, modern spatially resolved spectroscopic surveys have relied on visual classification of galaxy kinematic maps to identify dynamical families of galaxies.  This approach was pioneered by the SAURON and ATLAS$^{\rm 3D}$ surveys \citep[e.g.][]{Emsellem07,Emsellem11,Cappellari11b,Krajnovic11} and expanded to e.g. the MASSIVE Survey \citep[e.g.][]{Veale17}, the Sydney Australian astronomical observatory Multi-object Integral-field spectrograph Galaxy Survey \citep[SAMI, e.g.,][]{Cortese16,vandeSande17b,Foster18,vandeSande18}, the Mapping Nearby Galaxies at Apache Point Observatory \citep[MaNGA, e.g.][]{Greene17,Masters21,VazquezMata22} and simulations \citep[e.g.][]{Naab14,Lagos22}. 
As with morphology, computational and statistical tools have also been employed to efficiently sort galaxy kinematics into objective classes \citep[e.g.][]{Kalinova17,vandeSande21a,Sweet20}.

One may expect that certain kinematic features, especially faint ones or those on small spatial scales, may be difficult to identify in shallower or lower spatial resolution higher redshift data. Such features might include central $2\sigma$ (i.e. double maxima in the velocity dispersion map) or central kinematically decoupled cores (KDCs). Statistical samples of galaxies with sufficient spatial resolution spectroscopic data at intermediate redshifts are only just becoming available to make comparisons of kinematic visual classifications possible \citep[e.g.][]{Guerou17,Foster21,MunozLopez24}.

The Middle-Ages Galaxies Properties with Integral field spectroscopy (MAGPI) Survey\footnote{\url{http://magpisurvey.org/}} is a medium-deep Very Large Telescope (VLT) Multi-Unit Spectroscopic Explorer (MUSE) survey of 60 massive ($> 7 \times 10^{10}M_\odot$ ) central galaxies at redshifts 0.25 < z < 0.35 (primaries) and their immediate environment. The MAGPI sample was selected from the Galaxy and Mass Assembly (GAMA) survey \citep{Driver11} to span a broad range of environments \ halo masses (i.e. $11.35 \le \log ( M_{\rm halo}/M_\odot) \le 15.35$). This is achieved through dedicated observations of 56 x 1 arcmin$^2$ fields and supplemented with 4 massive centrals from archival observations of Abell 370 (Program ID 096.A-0710; PI: Bauer) and Abell 2744 (Program IDs: 095.A-0181 and 096.A-0496; PI: Richard)\footnote{Archival fields are not included in this work.}.
At the time that the visual classifications used herein were performed, only 35 of the 56 MAGPI fields had been completely observed and reduced. The remaining fields will be processed in due course.

The MAGPI survey is thus intermediate to data used in past visual classification studies. It is a relatively small survey (<1000 individual galaxies to classify), making careful visual inspection by a set of expert team members possible. However, MAGPI images and kinematic maps have respectively lower than and comparable spatial resolution to other $z\sim0$ comparable surveys such as MaNGA and SAMI. We thus expect that morphological classifications in MAGPI will be intermediate in detail and accuracy, lying between high redshift and local surveys, while kinematic classification will be of similar quality and accuracy to MaNGA and SAMI, but with the advantage of not being restricted to a fixed field-of-view/size.

In this work, we present morphological and dynamical visual classifications for all spatially resolved targets in 35 MAGPI fields.
We endeavour to identify and quantify the dynamical state of MAGPI galaxies and compare our visual dynamical classification with published results from local surveys. In \S\ref{sec:data}, we present the MAGPI data used in this work. \S\ref{sec:analysis} presents our statistical analysis of the collected visual classifications. Our results are outlined in \S\ref{sec:results} and discussed in \S\ref{sec:discussion}. We present our conclusions in \S\ref{sec:conclusions}.

We assume a $\Lambda$CDM universe with $H_0$ = 70 kms$^{-1}$ Mpc$^{-1}$, $\Omega_{\rm M}$ = 0.3 and $\Omega_\Lambda$ = 0.7.

\section{Data}\label{sec:data}

\subsection{Data reduction}

The data used herein are taken from the MAGPI Survey. Survey strategy, sample description and science goals are described in \citet{Foster21} along with a description of the initial data reduction. Detailed data reduction, curation and quality control will be discussed in Mendel et al. (in prep).

Briefly, raw MUSE data cubes are reduced using the ESO MUSE pipeline \citep{Weilbacher12,Weilbacher20} via the {\sc pymusepipe}\footnote{\url{https://github.com/emsellem/pymusepipe}} interface. Basic data reduction steps performed include bias and overscan subtraction, flat-fielding, wavelength calibration, telluric correction and cube reconstruction. Sky subtraction is further improved using the Zurich Atmosphere Purge \citep[ZAP,][]{Soto16} package.

Individual objects within the MAGPI cubes are detected on the white light image from the MUSE data cube using the {\sc ProFound r} package \citep{Robotham18}. {\sc ProFound} is used to define the edges (i.e. segmentation maps) for every source.

Reduced MUSE data cubes are then cut into ``minicubes'' following the individual segmentation maps for every object using {\sc mpdaf}\footnote{\url{https://github.com/musevlt/mpdaf}}. {\sc mpdaf} is further used to produce synthetic Sloan Digital Sky Survey (SDSS) filter images in $g$, $r$ and $i$. We note that because of the nominal MUSE wavelength and the notch filter that obscures the Na laser, the wavelength ranges of the $g$ and $r$-bands are only partially covered. These MAGPI synthetic images are used to produce RGB colour images for classification. The default image scaling from {\sc magicaxis/magimageRGB}\footnote{\url{https://github.com/asgr/magicaxis}} is used for visual classification and throughout this work.

\subsection{Galaxy property estimates}

\subsubsection{Redshifts}
For each detected object within the MAGPI field-of-view, a 1 arcsec aperture spectrum is created for each object. The finalised redshift is measured on this aperture spectrum using MARZ \citep{Hinton16} with visual inspection. We use a modified template set provided by M. Fossati \footnote{\url{https://matteofox.github.io/Marz/}}. This set includes templates with higher spectral resolution and out to higher redshifts, as well as a variety of source types representative of the range found in MAGPI.

\subsubsection{Structural parameters}
Basic structural parameters (e.g. effective radius $R_e$, photometric position angle $PA_{\rm phot}$, S\'ersic indices $n$, etc) in all three synthetic bands $g$, $r$ and $i$ are obtained using both {\sc ProFound} and {\sc GalFit} \citep{Peng02,Peng10}. Except for the initial selection described in \S\ref{sec:sample}, {\sc GalFit} measurements are used. {\sc ProFound} further provides integrated magnitudes in $g$, $r$ and $i$. 

\subsubsection{Stellar masses}
Stellar masses are derived in a method consistent to that used for the GAMA survey \citep{Bellstedt20, Driver22} using the {\sc ProSpect} SED-fitting code \citep{Robotham20}. We use images that are pixel-matched to the MAGPI minicubes from the Kilo-Degree Survey \citep[KiDS,][]{deJong17} and VISTA Kilo-degree Infrared Galaxy \citep[VIKING,][]{Edge13} in 9 bands ($u$-$K_s$) to apply forced photometry based on the MAGPI segmentation maps. We assume a skewed-normal star formation history with a linearly evolving metallicity, a \citet{Chabrier03} initial mass function (IMF) and use the \citet{Bruzual03} stellar population templates.

\subsection{Kinematic maps and parameters}

Stellar and ionised gas kinematic velocity ($v$) and dispersion ($\sigma$) maps are obtained independently using the penalised pixel fitting {\sc python} package ({\sc pPXF}, \citealt{Cappellari04,Cappellari17}) via the Galaxy Integral field unit Spectroscopy Tool {\sc GIST}\footnote{\url{https://pypi.org/project/gistPipeline/}; \url{https://github.com/geckos-survey/ngist}} (\citealt{Bittner19}; Fraser-McKelvie et al. in prep) as described in e.g., \citet{Foster21,Bagge23,DEugenio23a}; Battisti et al. (in prep).

Briefly, stellar kinematics are obtained by fitting the stellar continuum from individual spaxels with a signal-to-noise ratio above 3 per pixel after masking spectral regions of possible nebular emission and strong skylines using a series of stellar templates from the IndoUS stellar template library \citep{SanchezBlazquez06}. Our method is similar to that used for the SAMI Galaxy Survey \citep{vandeSande17a,Croom21}. Specifically, {\sc pPXF} is first fit to elliptical annular bins to determine an optimal subset of templates before individual spaxels are fit with subsets from their respective and adjacent radial bins (see \citealt{Foster21} and \citealt{DEugenio23a} for further detail). Only spaxels with valid values (i.e. excluding NaN or Inf) are presented for visual classification purposes. Stellar velocity and dispersion maps presented to classifiers were scaled between the 5$^{\rm th}$ and 95$^{\rm th}$ percentiles, but users are able to interactively modify those bounds and zoom in or out. This default scaling is used for plotting purposes throughout. Only classifications associated with stellar kinematic maps with a minimum of 15 valid spaxels are considered (231 galaxies).

Similarly, ionised gas kinematics are obtained through fitting and subtracting the stellar continuum using {\sc gist} and {\sc pPXF} with light-weighted stellar templates ssp\_mist\_c3k\_salpeter (Charlie Conroy, private communication) before using a customised version of {\sc pyGandALF} \citep{Sarzi06,FalconBarroso06} for emission line modelling. For visual classification purposes, we presents/plot only spaxels where the brightest emission line has a signal-to-noise ratio $> 3$. Ionised gas kinematic maps have been used and described by the MAGPI team in e.g. \citet{Foster21}, \citet{Bagge23} and \citet{Chen24}, and full detail of their production will be provided in Battisti et al. (in prep). As for the stellar kinematics, gas velocity and dispersion maps presented to classifiers were scaled between the 5$^{\rm th}$ and 95$^{\rm th}$ percentiles by default (also used for plotting purposes throughout this paper), with users able to interactive change those bounds and zoom in or out. Only classifications associated with ionised gas kinematic maps with a minimum of 15 valid spaxels are considered (402 galaxies).

For MAGPI data, we use the corrected spin parameter proxy values determined by \citet{Derkenne24}. Briefly, the spin proxy parameter $\lambda_{r}$ is computed based on the \textit{stellar} kinematic maps following the definition of \citet{Emsellem07}:
\begin{equation}
\lambda_r=\frac{\sum^{N}_{i=1} F_i R_i\abs{v_i}}{\sum^{N}_{i=1}F_i R_i \sqrt{v_i^2+\sigma_i^2}},
\end{equation}
where $F_i$, $R_i$, $v_i$ and $\sigma_i$ are the flux, galactocentric radius, recession velocity corrected for the systemic velocity and the velocity dispersion measured in the $i^{\rm th}$ spaxel within an aperture of size $r$, respectively. For our purposes, this quantity is measured within the $r=R_e$ elliptical aperture closely following the methodology of \citet[][]{FraserMcKelvie22}. Seeing and aperture corrections are applied following \citet{Harborne20}. More detail on the calculation of the corrected $\lambda_{R_e}$ values used herein can be found in \citet{Derkenne24}.

\subsection{Environment metrics}

Environmental metrics for MAGPI are calculated using {\sc parliament}\footnote{A commonly used collective noun for a group of magpies is a ``parliament''.} (Harborne et al. in prep) following the methodology of \citet{Knobel09} and \citet{Robotham11}. Based on galaxy groups identified with {\sc parliament}, we make use of the total number of group members ($N_{\rm group}$), the group mass proxy (assuming a multiplicative factor of $A=10$, see \citealt{Robotham11}) and the computed distance to the first nearest neighbour ($d_1$) as described in \citet{Bagge23}.

\section{Sample selection for classification}\label{sec:sample}
MAGPI team members were invited to visually inspect and classify galaxies from the MAGPI Master Catalogue (Mendel et al. in prep) that satisfy the following criteria:
\begin{enumerate}
\item $R_{\rm e, ProFound} > 0.75 \times$ FWHM$_r$ (i.e. source is extended in the $r$-band according to {\sc ProFound}),
\item $R_{\rm e, GalFit} > 0.75 \times$ FWHM$_r$ (i.e. source is extended in the $r$-band according to {\sc GalFit}), and
\item $i < 26$ mag (AB). 
\end{enumerate}
This selection is deliberate, yielding 637 galaxies that are potentially resolved across a redshift range of $0.08 \le z \le 5$ to visually classify (from 3970 MAGPI detections). We consider both $R_{\rm e, ProFound}$ and $R_{\rm e, GalFit}$ sizes for selection purposes to ensure that the majority of unresolved targets are removed. Only {\sc GalFit} values are used elsewhere in this work. Despite the above criteria and due to {\sc ProFound} and {\sc GalFit} sometimes failing in similar ways for targets near the edge of the MAGPI Fields where background noise dominates, a significant fraction of the faintest 637 objects are seemingly unresolved. These are small round objects with associated {\sc ProFound} and {\sc GalFit} sizes extending beyond their visible extent on the image.

\begin{figure*}
\includegraphics[width=18cm]{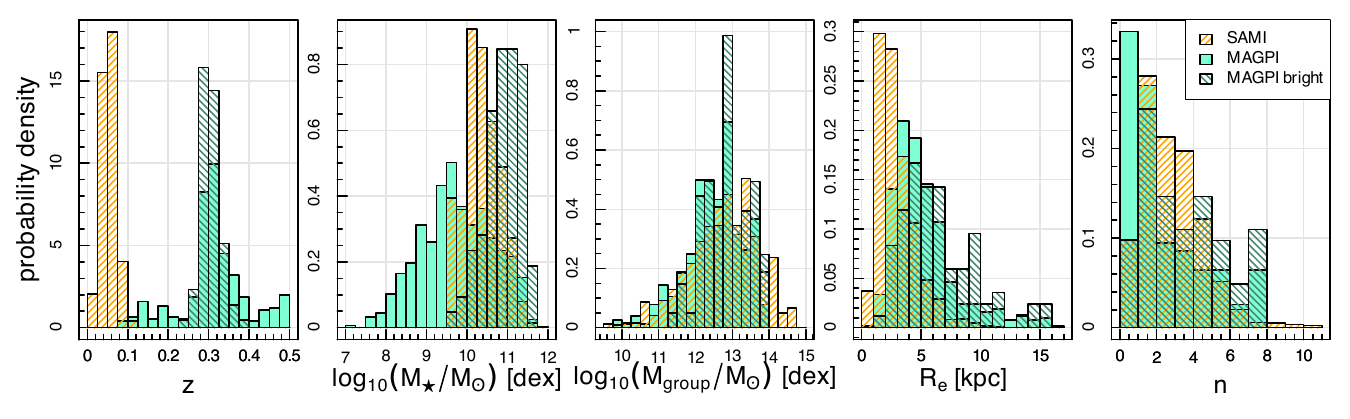}
\caption{\textbf{From left to right:} Redshift (z), stellar mass ($M_\star/M_\odot$), group mass proxy ($M_{\rm group}/M_\odot$), effective radius ($R_e$) and S\'ersic index (n) distributions for the visually classified (aquamarine filled histogram) and bright ($r<20$ and $0.2\le z \le 0.4$, green lined histogram) MAGPI samples. Distributions for the compared SAMI sample discussed in \S\ref{sec:discussion} are shown as orange lined histograms. The bright MAGPI sample is offset to higher stellar masses and effective radii than the SAMI galaxies. The ranges of S\'ersic indices and group masses probed are similar between the MAGPI bright and SAMI samples, suggesting a comparable mix of morphologies and environments.} \label{fig:SAMI_selection_comp}
\end{figure*}

We further compile and discuss a `bright sample' of 86 intermediate redshift galaxies ($0.2\le z \le0.4$) with $r<20$ mag. Classifications for galaxies fainter than this threshold appreciably increase in difficulty and decrease in reliability (refer to \S\ref{sec:analysis}, \ref{sec:appendix_threshold}). Stellar mass, effective radius and S\'ersic index histograms for the visually classified and bright MAGPI samples are shown in Figure \ref{fig:SAMI_selection_comp}.

\subsection{Collection of independent opinions}\label{sec:collection}

\begin{table*}
\caption{Summary of questions (Column 1) and possible user input (Column 4) included in the {\sc r-Shiny} web application$^{\rm a}$. Parameter and corresponding numerical values are given in Columns 2 and 3, respectively. Column 5 lists relevant instructions to classifiers provided within the application.}
\begin{tabular}{ |l|c| c c l| }
\hline
\textbf{Question} & \textbf{Parameter} & \textbf{Numeric Value} & \textbf{Corresponding Option} & \textbf{Relevant instructions}\\
\textbf{(1)} & \textbf{(2)} & \textbf{(3)} & \textbf{(4)} & \textbf{(5)}\\
\hline
\hline
\textbf{Morphological classification} & Morph &  0 & None selected (NS)     & \\\cline{3-5}
                             &       &  1 & Elliptical (E)     & smooth, featureless\\\cline{3-5}
                             &       &  2 & Lenticular (S0)    & obvious disk, no evidence for spiral \\
                             &       &    &                    & arms, substantial bulge\\\cline{3-5}
                             &       &  3 & early Spiral (eSp) & evidence for spiral arms and a bulge\\\cline{3-5}
                             &       &  4 & late Spiral (lSp)  & prominent spiral arms and minimal/\\
                             &       &    &                    & no bulge\\\cline{3-5}
                             &       &  5 & Irregular (Irr)    & no distinct regular shape\\\cline{3-5}
                             &       &  6 & Merger (Mer)       & visual evidence of ongoing merger\\\cline{3-5}
                             &       &  7 & I don't know (IDK) & \\
\hline
\textbf{This galaxy is barred}        & BarFlag &  0 & Not selected & \\\cline{3-5}
                             &         &  1 & Selected     & the galaxy has an evident bar\\
\hline
\textbf{There are other features in the image} & VisFeatFlag &  0 & Not selected & \\\cline{3-5}
                                      &             &  1 & Selected     & unlisted features in the image\\
\hline
\hline
\textbf{Stellar kinematics classification} & StellOR    & 0 & None selected (NS)                  & \\\cline{3-5}
                                  &            & 1 & Obvious Rotation (OR)           & clear rotation, possibly accompanied\\
                                  &            &   &                                 & by peaked or flat sigma\\\cline{3-5}
                                  &            & 2 & No obvious Rotation (NOR)       & \\\cline{3-5}
                                  &            & 3 & I don't know (IDK)              & \\
\hline
\textbf{Stellar kinematics features}       & StellFeat  & 0 & None selected (NS)                  & \\\cline{3-5}
                                  &            & 1 & Without feature(s) (WOF)        & \\\cline{3-5}
                                  &            & 2 & With feature(s) (WF)            & unusual features in the kinematic\\
                                  &            &   &                                 & maps such as 2 velocity dispersion\\
                                  &            &   &                                 & peaks, changes in the kinematic\\
                                  &            &   &                                 & position angle such as kinematically\\
                                  &            &   &                                 & decouple cores (KDC) or kinematic\\
                                  &            &   &                                 & twists (KT)\\\cline{3-5}
                                  &            & 3 & I don't know (IDK)              & \\
\hline
\textbf{There are issue(s) with the stellar}                & StellKinFlag & 0 & Not selected & \\\cline{3-5}
\textbf{kinematic maps}                                     &              & 1 & Selected     & issues may limit stellar kinematic data's\\
                                                   &              &   &              & usability for science\\
\hline
\hline
\textbf{Gas kinematics classification} & GasOR      & 0 & None selected (NS)                      & \\\cline{3-5}
                              &            & 1 & Obvious Rotation (OR)               & clear rotation, possibly accompanied\\
                              &            &   &                                     & by peaked or flat sigma\\\cline{3-5}
                              &            & 2 & No obvious Rotation (NOR)           & \\\cline{3-5}
                              &            & 3 & I don't know (IDK)                  & \\
\hline
\textbf{Gas kinematic features}        & GasFeat    & 0 & None selected (NS)                      & \\\cline{3-5}
                              &            & 1 & Without feature(s) (WOF)            & \\\cline{3-5}
                              &            & 2 & With feature(s) (WF)                & unusual features in the kinematic\\
                              &            &   &                                     & maps such as 2 velocity dispersion\\
                              &            &   &                                     & peaks, changes in the kinematic\\
                              &            &   &                                     & position angle such as kinematically\\
                              &            &   &                                     & decouple cores (KDC) or kinematic\\
                              &            &   &                                     & twists (KT)\\\cline{3-5}
                              &            & 3 & I don't know (IDK)                  & \\
\hline
\textbf{There are issue(s) with the gas}  & GasKinFlag & 0 & Not selected    & \\\cline{3-5}
\textbf{kinematic maps}                   &              & 1 & Selected     &  issues may limit gas kinematic data's\\
                                 &              &   &              & usability for science\\
\hline
\end{tabular}
\label{table:app_options}
\end{table*}

We use a bespoke online application to collect individual and independent classifications. Each classifier receives their own randomised list of galaxies to classify. As a result, although there were 15 classifiers, each galaxy was independently assessed by 11-13 individuals.

The {\sc r} code for the visual classification web application is publicly available on \href{https://github.com/MagpiSurvey/MAGPIClassApp_public}{GitHub}\footnote{\url{https://github.com/MagpiSurvey/MAGPIClassApp_public}}.
The \href{https://shiny.datacentral.org.au/magpiclassapp/}{MAGPI visual classification web application} was deployed on \href{https://datacentral.org.au}{Data Central}'s {\sc shiny}\footnote{\url{https://www.rstudio.com/products/shiny/}} server.

\begin{figure*}
\centering
\includegraphics[width=18cm]{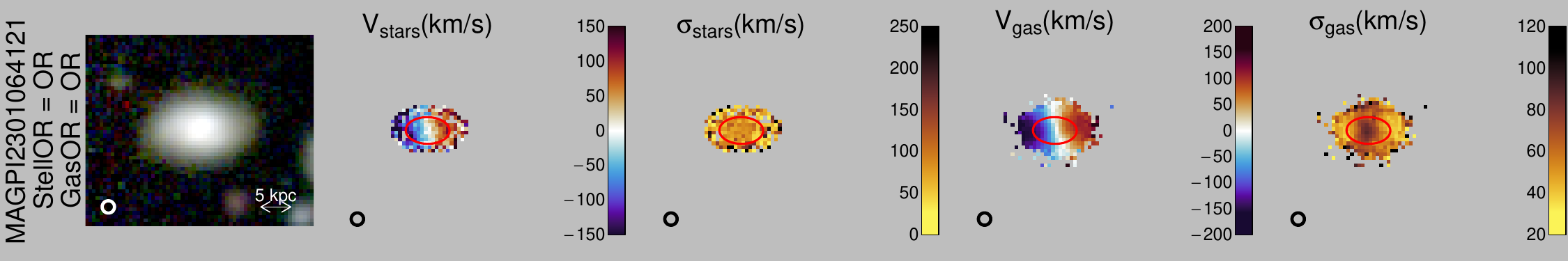}\\
\includegraphics[width=18cm]{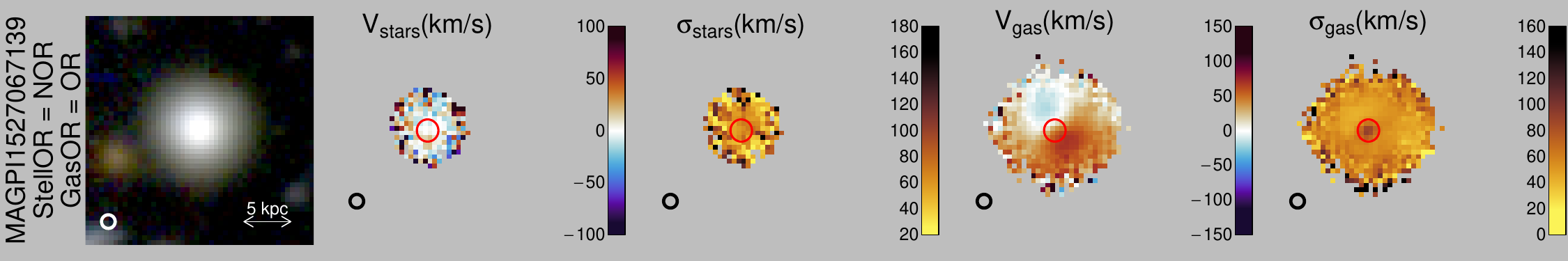}\\
\includegraphics[width=18cm]{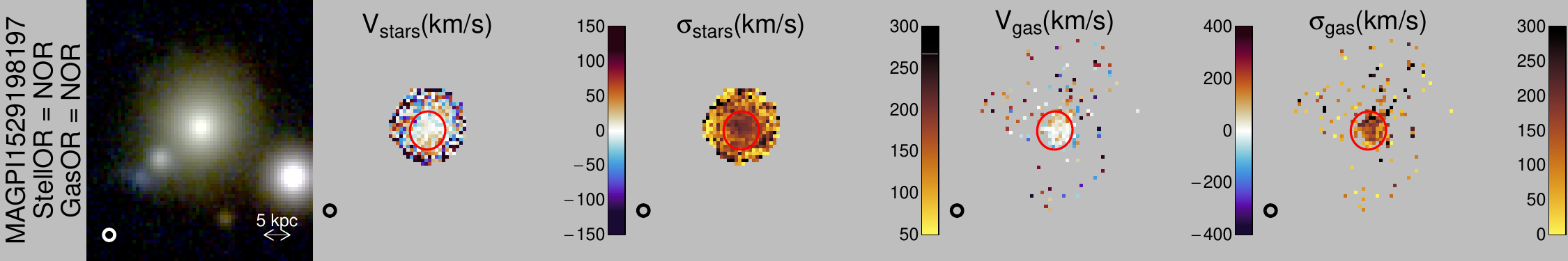}
\caption{Example MAGPI galaxies in each of the rotational classes. From top to bottom as labelled on the left: MAGPI2301064121 with obvious stellar and gas rotation (i.e., Stars OR, Gas OR); MAGPI1527067139 with non-obvious stellar and obvious gas rotation (i.e., Stars NOR, Gas OR); and MAGPI1529198197 with non-obvious stellar and gas rotation (i.e., Stars NOR, Gas NOR). There is no galaxy with reliable Stars OR and Gas NOR. From left to right: synthetic $g$, $r$, $i$-band colour image of the galaxy based on the MUSE data; stellar velocity map; stellar dispersion map; gas velocity map; and gas dispersion map. The PSF is shown as a white or black circle in the bottom-left corner of each panel. All panels within a row are on the same scale, and an arrow representing the physical scale (in kpc) is shown on the left-most panel for each galaxy. Red ellipses represent $1R_e$.}\label{fig:Kin_Gallery}
\end{figure*}

Briefly, volunteers from the MAGPI team (henceforth ``classifiers'') were asked to give their opinion based on the MAGPI synthetic colour image, stellar and gas kinematic (velocity and dispersion) maps. Example images and kinematic maps presented to classifiers are shown in Figure \ref{fig:Kin_Gallery}. Relevant information regarding the questions and options provided to classifiers are summarised in Table \ref{table:app_options}. These questions reflect the classifications used in SAMI \citep{vandeSande21a}.

\begin{figure}
\includegraphics[width=8.5cm]{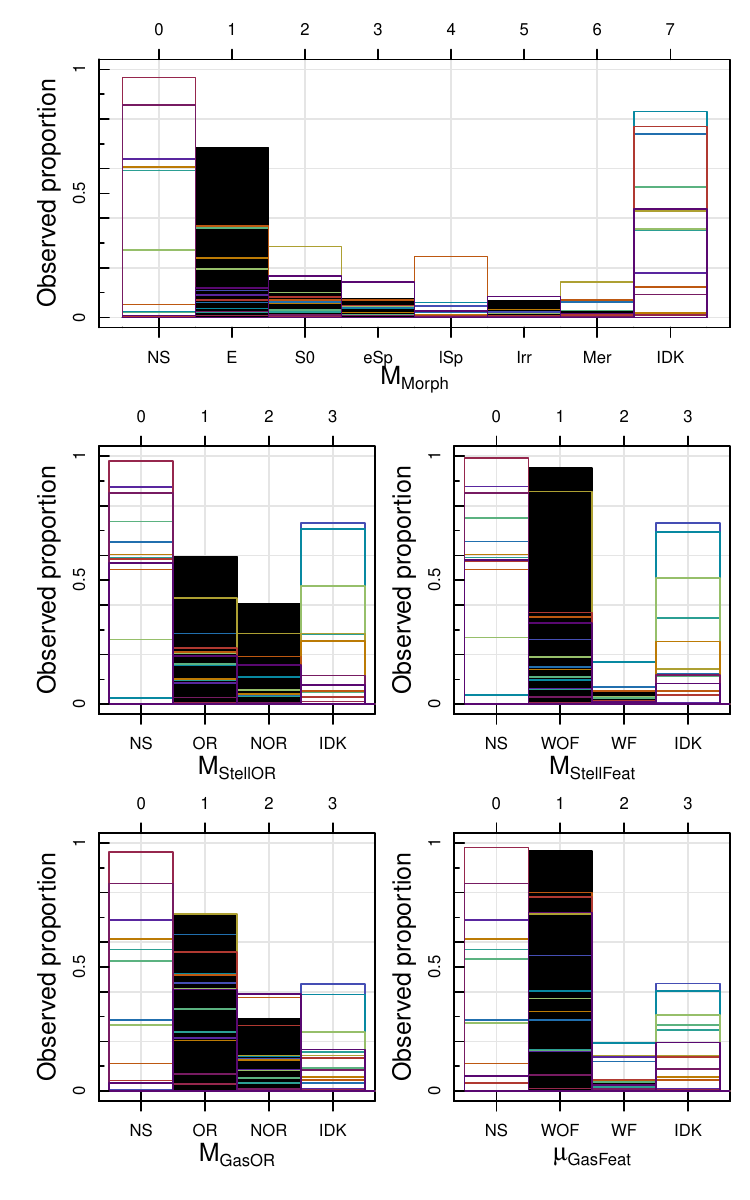}
\caption{Posterior mode ($\mu$) observed proportions (black histograms) with superimposed raw observed proportions for individual classifiers shown as different coloured histograms prior to applying quality cuts. These histograms illustrate the range of responses received as input for our Bayesian modelling. Parameters are as per Table \ref{table:app_options} for morphology (Morph, top row), stellar kinematic (StellOR and StellFeat, middle row) and gas kinematics (GasOR and GasFeat, bottom row). There are broad differences in the shape of the distributions from classifier to classifier. The individual distributions also illustrate how different classifiers favoured ``IDK'' or ``NS'' and the high proportion of each features that were difficult to classify.}\label{fig:classifiers_compared}
\end{figure}

 Visual classifications were performed on the team data for the 35 science-ready MAGPI fields. A compilation of the raw output from individual visual classifications is presented in Figure \ref{fig:classifiers_compared}, showing the range of opinions across the sample. The distribution of responses varied substantially between classifiers, with some exhibiting more confidence in assigning a category than others, who conversely prefer either not to select a category (NS) or explicitly state they do not know (IDK). \S\ref{sec:analysis} describes how these categories are treated within our Bayesian inference. Given that all classifiers are astronomers (graduate astronomy students and beyond), we assume that all approaches are equally valid and that this diversity of opinions reflects the inherent subjectivity of visual classifications. Therefore, in what follows, we assume that there is no individual whose opinion holds the ``ground truth'' and that the decision of the majority should be preferred over that of individuals. Indeed, we find only very weak and incoherent trends (not shown here for privacy reasons) between the year of a classifier's first publication (as a proxy for reverse expertise) and their willingness to assign a visual classification.

\section{Analysis}\label{sec:analysis}

For each galaxy, visual classifications for each collected parameter (i.e., Morph, StellOR, StellFeat, GasOR, and GasFeat) are aggregated as follows. First the modes ($M$) of the relevant distributions are computed after excluding NS (none selected). We exclude from subsequent analysis classifications based on stellar (399 galaxies) or gas (122 galaxies) kinematic maps where 3 or more classifiers raised the “there are issue(s)" flags (i.e. StellKinFlag or GasKinFlag, respectively).

Statistical inferences on galaxy parameters are then derived using a modified Dawid--Skene model \citep{Dawid79}, extended to allow for NS and IDK responses. The standard Dawid--Skene model posits that, given a galaxy with true parameter $z$, the classification provided by classifier $i$ is drawn from a categorical distribution with probability vector $\boldsymbol{\theta}_{iz}$, which represent classifier-specific response tendencies for galaxies of that type. By considering the concordance and discordance between classifiers' classifications of the same galaxies, and assuming that the classifiers are generally accurate (i.e. high probability of providing correct classifications), the classifier tendencies and true parameters that are most compatible with the observed pattern of classifications can be identified.

Our extended model---the full details of which are provided in \ref{sec:appendix_dsmodel}---introduces classifier-specific confidence parameters and galaxy-specific difficulty parameters, which influence the rate at which classifiers respond NS or IDK. Further, the model assumes that classifier accuracy (the probability of providing the correct classification) is lower for galaxies with higher difficulty.

Bayesian inference is performed by modifying the {\sc rater} package \citep{Pullin23} to implement the extended model in {\sc RStan} \citep[version 2.32.6;][]{RStan}, which uses a dynamic Hamiltonian Monte Carlo algorithm to estimate the posterior distribution. Weakly informative priors are used for all parameters. In particular, for classifier tendencies $\boldsymbol{\theta}$ related to a parameter with $K$ possible values, we use a Dirichlet prior with concentration parameter $7(K-1)/6$ on the correct classification and $1/2$ on each incorrect classification, such that the probability of a correct classification for each true parameter has an expected value of 70\% but could range from 8\% to 99.9\% when $K = 2$, or from 37\% to 94\% when $K = 6$, with 95\% prior probability.

\begin{figure}
\includegraphics[width=8.5cm]{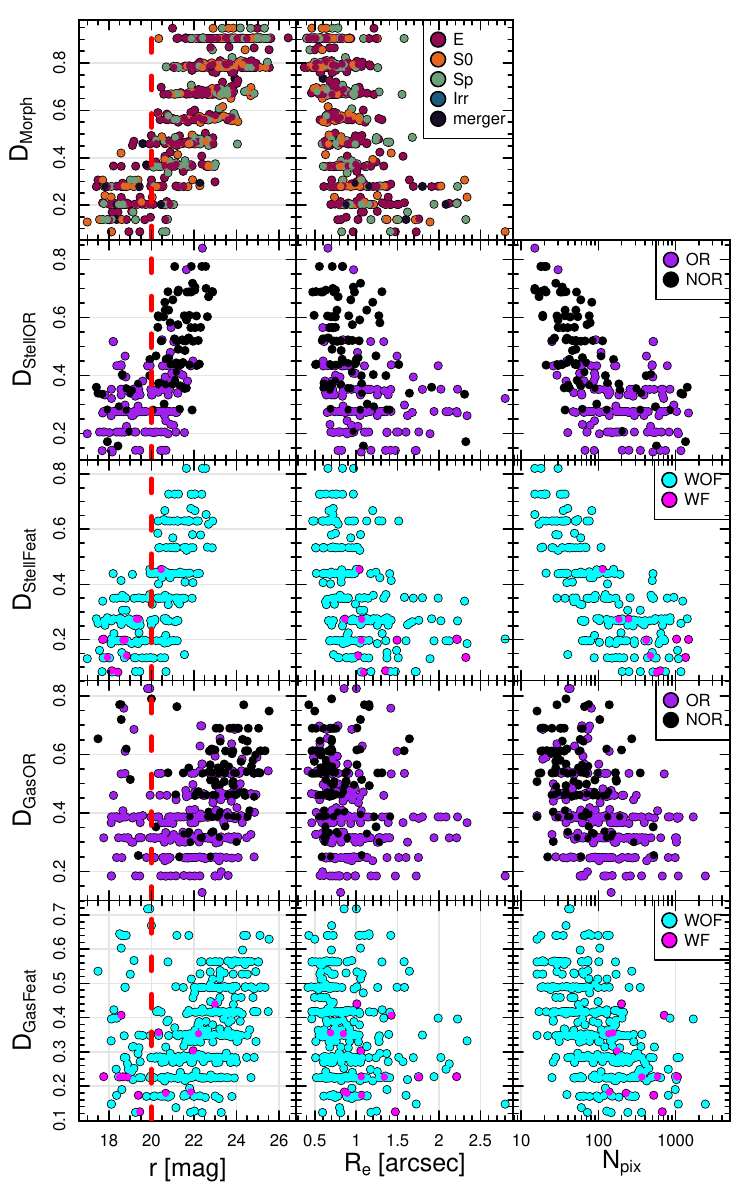}
\caption{Dependence of the ``difficulty'' parameter for morphology ($D_{\rm Morph}$), stellar rotation ($D_{\rm StellOR}$), stellar features ($D_{\rm StellFeat}$), ionised gas rotation ($D_{\rm GasOR}$) and features ($D_{\rm GasFeat}$) with $r$-band magnitude (left), effective radius (middle), and the number of good pixels in the relevant map shown to classifiers ($N_{\rm pix}$). Points are colour-coded according to the posterior mode of each galaxy as per the legend shown on the right-most panel of each row. Red dashed lines show the threshold for the ``bright sample'' at $r=20$ mag. Because stellar kinematics require comparatively higher surface brightnesses than images and gas kinematics, stellar kinematic maps are not available to visually classify galaxies fainter than $r\sim23$, explaining the shorter range of magnitudes covered by StellOR and StellFeat. In general, fainter and smaller galaxies are more difficult to classify. Galaxies with obvious rotation (StellOR and GasOR) and dynamical features (StellFeat and GasFeat) are easier to classify (i.e. lower mean/median $D$).}\label{fig:difficulties}
\end{figure}

For each galaxy and for each parameter, we obtain the vector of posterior probabilities $P_k$ that the underlying truth corresponds to each option $k$. The largest of these defines the posterior mode, denoted $\mu_{\rm Morph}$, $\mu_{\rm StellOR}$, $\mu_{\rm StellFeat}$, $\mu_{\rm GasOR}$ and $\mu_{\rm GasFeat}$, respectively. We also extract the posterior mean of the difficulty of classifying each parameter for each galaxy, denoted $D_{\rm Morph}$, $D_{\rm StellOR}$, $D_{\rm StellFeat}$, $D_{\rm GasOR}$ and $D_{\rm GasFeat}$, and representing the expected probability of a IDK or NS response from the average classifier. There were no explicit instructions on how to treat IDK vs. NS classifications. Given that individual choices on how to treat those options differed across raters and that neither response provides additional information, we decided not to distinguish them in the model and instead interpret both answers as reflecting hesitation in rating a given property. Figure \ref{fig:difficulties} shows how the difficulty of classifying the different features of a galaxy strongly depends galaxy brightness, size and the number of available pixels (for the kinematic maps).

\begin{figure}
\centering
\includegraphics[width=9cm]{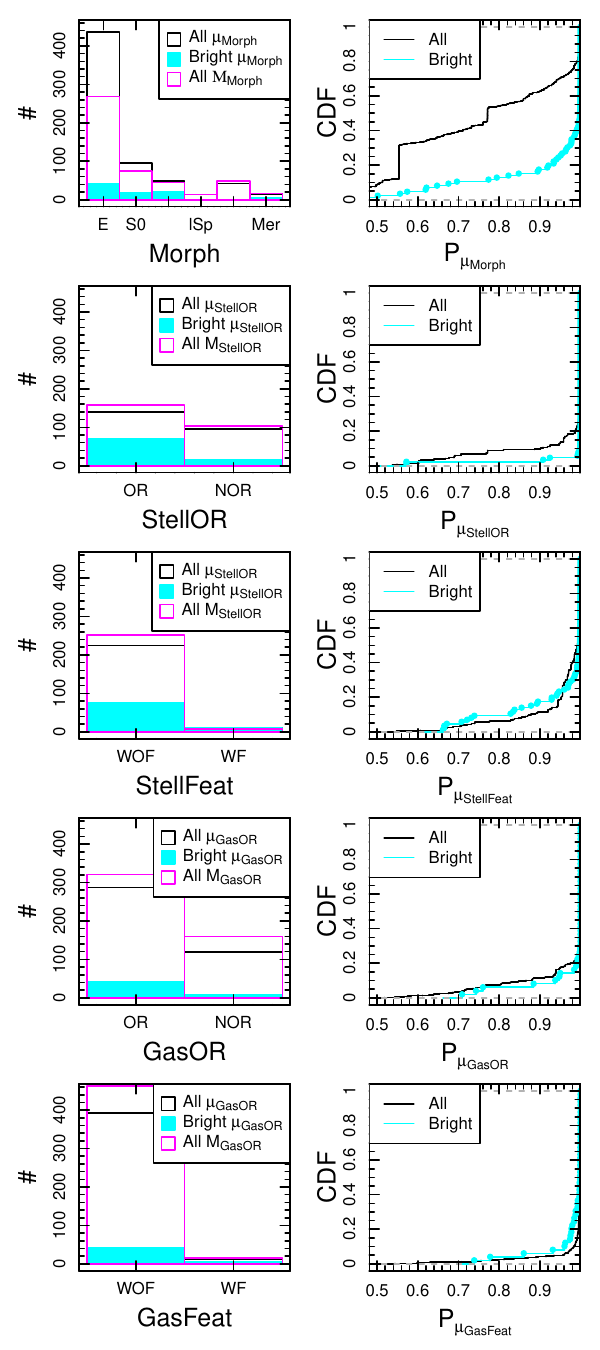}
\caption{Comparison of the output distributions (left) for the Bayesian and simple consensus approaches for (from top to bottom) Morph, StellOR, StellFeat, GasOR and GasFeat. Raw ($M$, magenta) and posterior ($\mu$) modes for the whole (black) and bright ($r<20$, cyan) samples as per the legend. Empirical cumulative distribution functions (CDF, right) for the posterior mode probability $P_{\mu_{\rm Morph}}$ for the whole (black) and bright (cyan) samples as per the legend.}\label{fig:Mode_hist}
\end{figure}

Figure \ref{fig:Mode_hist} compares the distributions of the Bayesian posterior modes ($\mu$) with that of the more commonly used simple mode ($M$) for each parameter. The cumulative distribution function of the posterior probabilities for the bright and whole samples are also shown. For most parameters, the bright sample of galaxies have a higher proportion of reliable ($P_\mu>0.98$) classifications.

Examples of the most uncertain visual morphological and dynamical classifications ($P_\mu < 0.55$) are shown in \ref{sec:appendix_threshold}. Uncertain morphologies are usually associated with galaxies of intermediate morphological types or merging systems, uncertain stellar dynamics are usually associated with complex kinematic maps, while uncertain ionised gas dynamical classifications are associated with sparsely sampled and/or noisy maps. For most science cases, we recommend such low $P_\mu$ classifications be removed.

Example galaxies spanning the range of StellOR, StellFeat, StellKinFlag, GasOR and GasFeat categories are shown in Figure \ref{fig:Kin_Gallery}. 

Finally, we compute the fraction of classifiers who have seen the relevant map and raised specific flags (i.e. BarFlag, VisFeatFlag, StellKinFlag and GasKinFlag):
\begin{equation}
    f_{\rm BarFlag} = \frac{N_{{\rm BarFlag = 1,\ Morph} \neq {\rm\ NS}}}{N_{\rm Morph \neq NS}},
\end{equation}
\begin{equation}
    f_{\rm VisFeatFlag} = \frac{N_{{\rm VisFeatFlag = 1,\ Morph} \neq {\rm\ NS}}}{N_{\rm Morph \neq NS}},
\end{equation}
\begin{equation}
    f_{\rm StellKinFlag} = \frac{N_{{\rm StellKinFlag = 1,\ StellOR} \neq {\rm\ NS}}}{N_{\rm StellOR \neq NS}},
\end{equation}
\begin{equation}
    f_{\rm GasKinFlag} = \frac{N_{{\rm GasKinFlag = 1,\ GasOR} \neq {\rm\ NS}}}{N_{\rm GasOR \neq NS}}
\end{equation}
These may be used to determine samples of galaxies where visual signs of a bar ($f_{\rm BarFlag} \ne0$), visual features/issues on the galaxy image ($f_{\rm VisFeatFlag} \ne0$) or issues with the kinematic maps ($f_{\rm StellKinFlag} \ne0$ or $f_{\rm GasKinFlag} \ne0$) have been raised. Example barred galaxies ($f_{\rm BarFlag} \ne0$) are shown in Figure \ref{fig:Tuning_Fork}.

\begin{table}[!h]
\begin{center}
\begin{tabular}{| l  c  c|}
\hline
\textbf{Category} & \textbf{$N$} & \textbf{\textit{f}}\\
(1) & (2) & (3)\\
\hline\hline
$r<20$ mag, $0.2\le z \le 0.4$ (bright sample) & 86 & 1\\
\hline
\textit{Morphology} &&\\
\hline
$P_{\mu_{\rm Morph}} > 0.98$ & 56 & $56/86=0.65$ \\
$\mu_{\rm Morph} =$ E & 29 & $29/56=0.52$ \\
$\mu_{\rm Morph} =$ S0 & 6 & $6/56=0.11$ \\
$\mu_{\rm Morph} =$ Sp & 16 & $16/56=0.29$ \\
$\mu_{\rm Morph} =$ Irr & 0 & $0/56=0$ \\
$\mu_{\rm Morph} =$ Mer & 5 & $5/56=0.089$ \\
\hline
\textit{Obvious rotation} &&\\
\hline
$P_{\mu_{\rm StellOR}} > 0.98$ & 82 & $82/86=0.95$ \\
$\mu_{\rm StellOR} =$ OR & 70 & $70/82=0.85$ \\
$\mu_{\rm StellOR} =$ NOR & 12 & $12/82=0.15$ \\
\hline
$P_{\mu_{\rm GasOR}} > 0.98$ & 42 & $42/86=0.49$ \\
$\mu_{\rm GasOR} =$ OR & 38 & $38/42=0.9$ \\
$\mu_{\rm GasOR} =$ NOR & 4 & $4/42=0.095$ \\
\hline
($P_{\mu_{\rm StellOR}} > 0.98$) \& ($P_{\mu_{\rm GasOR}} > 0.98$) & 39 & $39/86=0.45$ \\
($\mu_{\rm StellOR} = $ OR) \& ($\mu_{\rm GasOR} = $ OR) & 34 & $34/39=0.87$ \\
($\mu{\rm StellOR} = $ OR) \& ($\mu_{\rm GasOR} = $ NOR) & 0 & $0/39=0$ \\
($\mu_{\rm StellOR} = $ NOR) \& ($\mu_{\rm GasOR} = $ OR) & 1 & $1/39=0.026$ \\
($\mu_{\rm StellOR} = $ NOR) \& ($\mu_{\rm GasOR} = $ NOR) & 4 & $4/39=0.1$ \\
\hline
\textit{Kinematic Features} &&\\
\hline
$P_{\mu_{\rm StellFeat}} > 0.98$ & 61 & $61/86=0.71$ \\
$\mu_{\rm StellFeat} = $ WOF & 56 & $56/61=0.92$ \\
$\mu_{\rm StellFeat} = $ WF & 5 & $5/61=0.082$ \\
\hline
$P_{\mu_{\rm GasFeat}} > 0.98$ & 38 & $38/86=0.44$ \\
$\mu_{\rm GasFeat} = $ WOF & 34 & $34/38=0.89$ \\
$\mu_{\rm GasFeat} = $ WF & 4 & $4/38=0.11$ \\
\hline
($P_{\mu_{\rm StellFeat}} > 0.98$) \& ($P_{\mu_{\rm GasFeat}} > 0.98$)& 32 & $32/86=0.37$ \\
($\mu_{\rm StellFeat} = $ WF) \& ($\mu_{\rm GasFeat} = $ WF) &2 & $2/32=0.062$ \\
\hline
\end{tabular}

\caption{Number ($N$, Column 2) and fraction ($f$, Column 3) of galaxies with $r<20$ mag (bright sample) and relevant $P_{\mu_{\rm Class}}>0.98$ in each morphological and kinematic category (Column 1). In each category, the proportion is calculated against the number of galaxies with comparable classification.}\label{table:families}
\end{center}
\end{table}

Free-form comments are also collated, although those are not used in this work.

In what follows, we refer to and generally consider ``reliable'' classifications as those with posterior mode probabilities $P_{\mu}>0.98$, indicating we ascribe less than a 2 percent chance that the feature is wrongly classified. 

\section{Results}\label{sec:results}

Basic descriptive statistics for galaxies in each category are summarised in Table \ref{table:families}. A brief summary of salient outcomes is given below.

\begin{figure*}
\centering
\includegraphics[width=16cm]{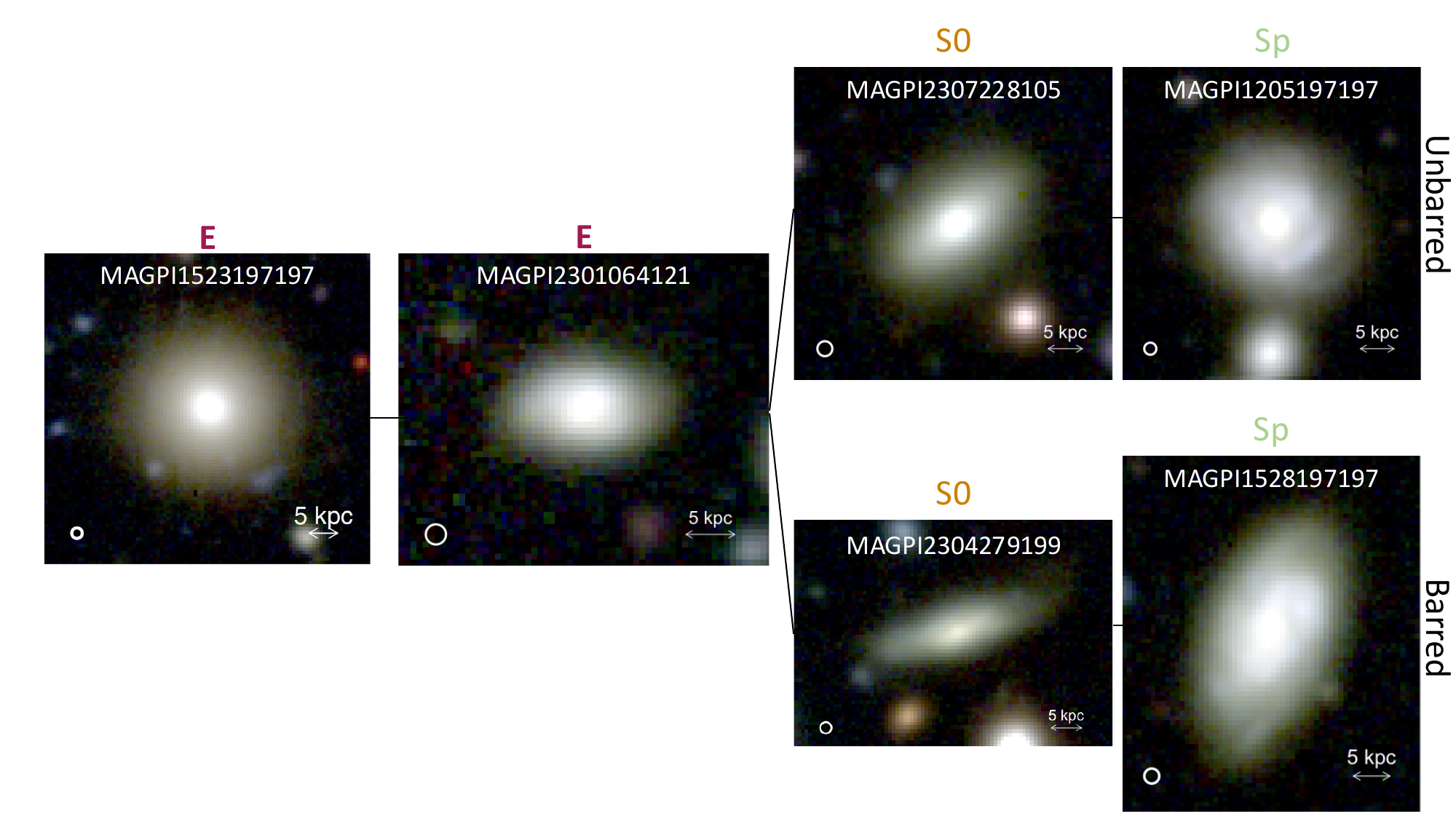}
\caption{The $z\sim0.3$ Hubble Tuning Fork using example MAGPI synthetic $g$, $r$, $i$-band colour images.}\label{fig:Tuning_Fork}
\end{figure*}

%Morphology
Figure \ref{fig:Morph_Gallery} shows example galaxies in each morphology category. Very few classifiers selected the Morph $=4$ (late spiral/lSp, see Figure \ref{fig:classifiers_compared}) overall, suggesting that late and early spirals are difficult to delineate in MAGPI-resolution images. We thus henceforth combine the lSp and eSp categories into a single Sp category. Our sample spans a broad range of morphologies (see Figures \ref{fig:Mode_hist}, \ref{fig:Tuning_Fork} and \ref{fig:Morph_Gallery}), confirming that the range of morphological classes is represented in MAGPI. Out of the 86 bright sample galaxies within our sample, 56 have reliable morphologies (i.e. $P_{\mu_{\rm Morph}}>0.98$), of which 52, 11, 29 and 5 percent are classified as E, S0, Sp and Mergers, respectively. While 5 merging galaxies are identified, there is no irregular galaxy in our bright sample.

%Rotation
Similarly, of the 82 (95 percent of the bright sample) galaxies with a reliably assigned stellar kinematic classification ($P_{\mu_{\rm StellOR}}>0.98$), 85 percent show obvious rotation (see Table \ref{table:families}). For the gas kinematics, this fraction is 90 percent, with 87 percent of galaxies showing obvious rotation in both maps. In other words, most galaxies in our sample rotate and when both gas and stars are present, both tend to rotate. There is however one exception where the stars do not show obvious rotation while the gas does (MAGPI1527067139), which is shown in the second row of Figure \ref{fig:Kin_Gallery}. There are only four galaxies (MAGPI1206110186, MAGPI1503197197, MAGPI1507084083 and MAGPI1529198197) for which there is no obvious rotation in either the stars or the gas (the latter is shown in Figure \ref{fig:Kin_Gallery}).

%Features
Of the 61 (38) bright galaxies with reliable stellar (gas) kinematic feature classifications, 5 (4) galaxies have $\mu_{\rm StellFeat}=$WF ($\mu_{\rm GasFeat}=$WF) (see Table \ref{table:families}). Kinematic features are proportionally slightly more common in the ionised gas kinematic maps and coexist in only 2 galaxies (MAGPI2310167176 and MAGPI2310199196). Example galaxies with stellar and gas kinematic features are shown in Figures \ref{fig:Stell_Feat} and \ref{fig:Gas_Feat}, respectively. Most show evidence of ongoing interaction, suggesting complex dynamics in MAGPI are usually associated with galaxy-galaxy interactions. As well as signs of interactions, a transition in the amplitude of rotation is sometimes seen in galaxies with $\mu_{\rm StellFeat}=$WF.

\begin{figure*}
\centering
\includegraphics[width=4.5cm]{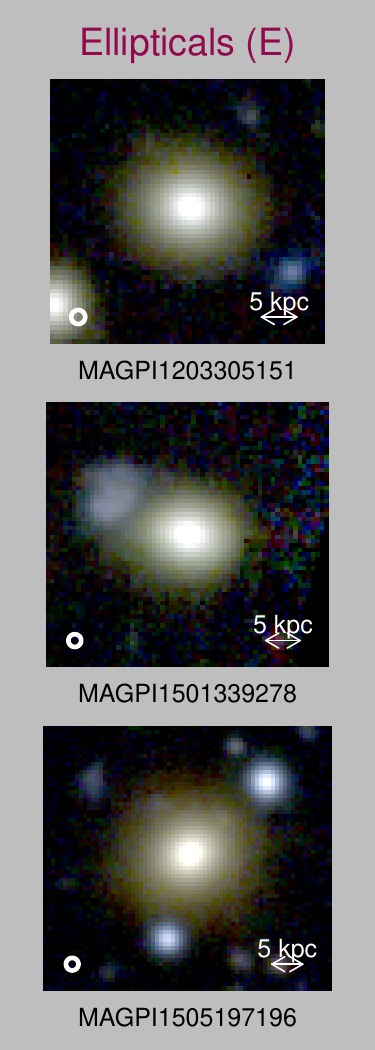}\includegraphics[width=4.5cm]{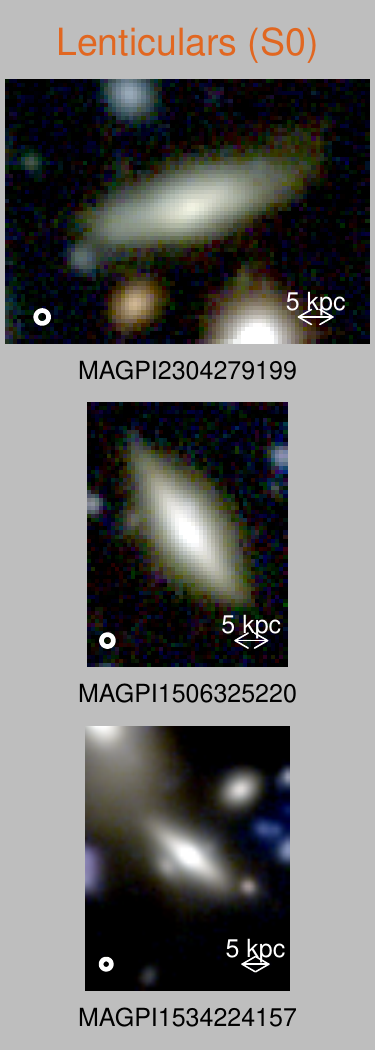}\includegraphics[width=4.5cm]{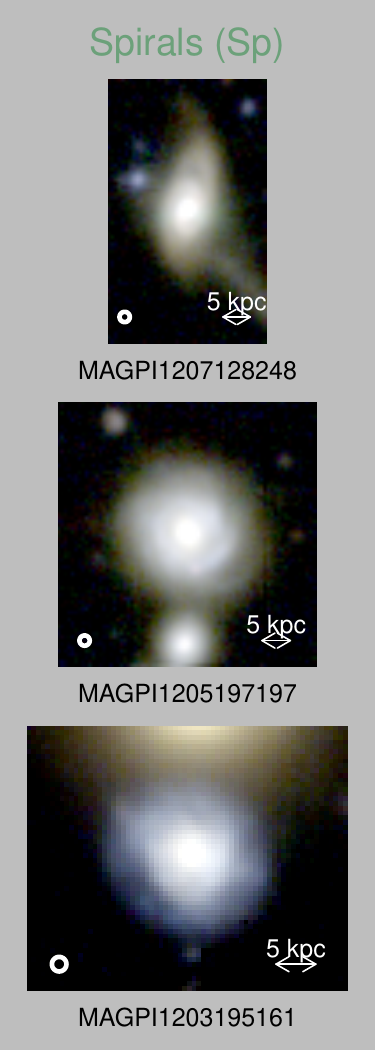}\includegraphics[width=4.5cm]{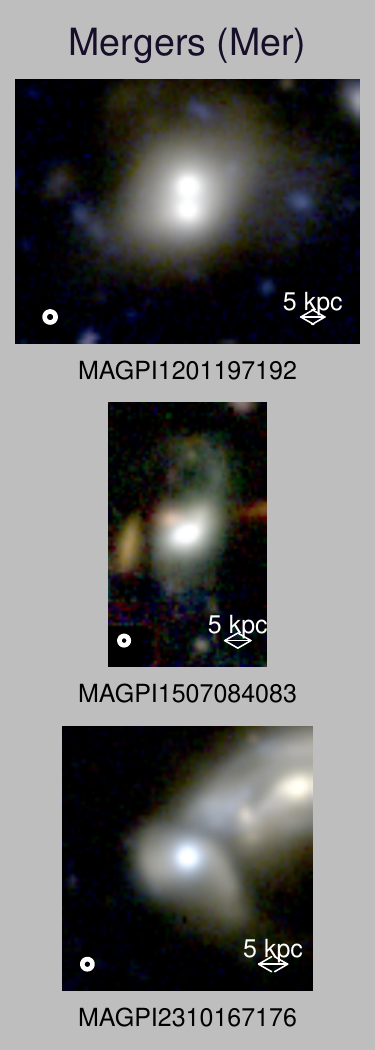}
\caption{Gallery of MAGPI synthetic $g$, $r$, $i$-band colour images of a selection of bright ($r<20$ mag) MAGPI galaxies with reliable visual morphologies. From left to right: example elliptical (E), lenticulars (S0), Spirals (eSp or lSp) and mergers (Mer). The PSF is shown as a white circle and the physical scale (in kpc) is represented as a white arrow on in each panel for reference.}\label{fig:Morph_Gallery}
\end{figure*}

\begin{figure}
\centering
\includegraphics[width=8.5cm]{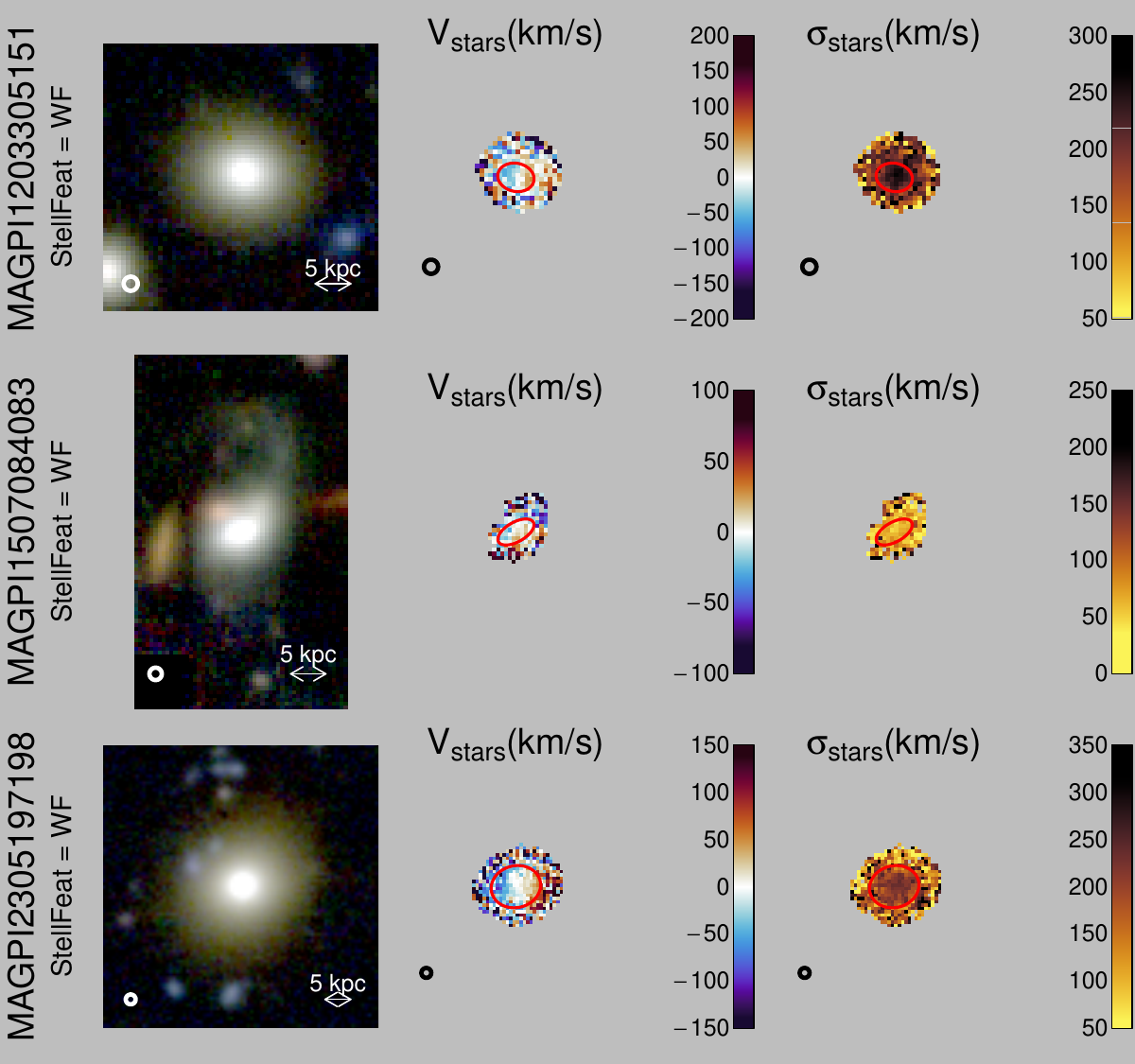}
\caption{Example MAGPI galaxies with stellar kinematic feature(s). From top to bottom as labelled on the left: MAGPI1203305151 (E with radial change in rotation); MAGPI1507084083 (Merger with complex velocity field); and MAGPI2305197198 (E with radial change in rotation amplitude). From left to right: MAGPI synthetic $g$, $r$, $i$-band colour image of the MAGPI target based on the MUSE data; stellar velocity map; and stellar dispersion map. The PSF is shown as a white or black circle in the bottom-left corner of each panel. All panels within a row are on the same scale, and an arrow representing the physical scale (in kpc) is shown on the left-most panel for each galaxy. Red ellipses represent $1R_e$.}\label{fig:Stell_Feat}
\end{figure}

\begin{figure}
\centering
\includegraphics[width=8.5cm]{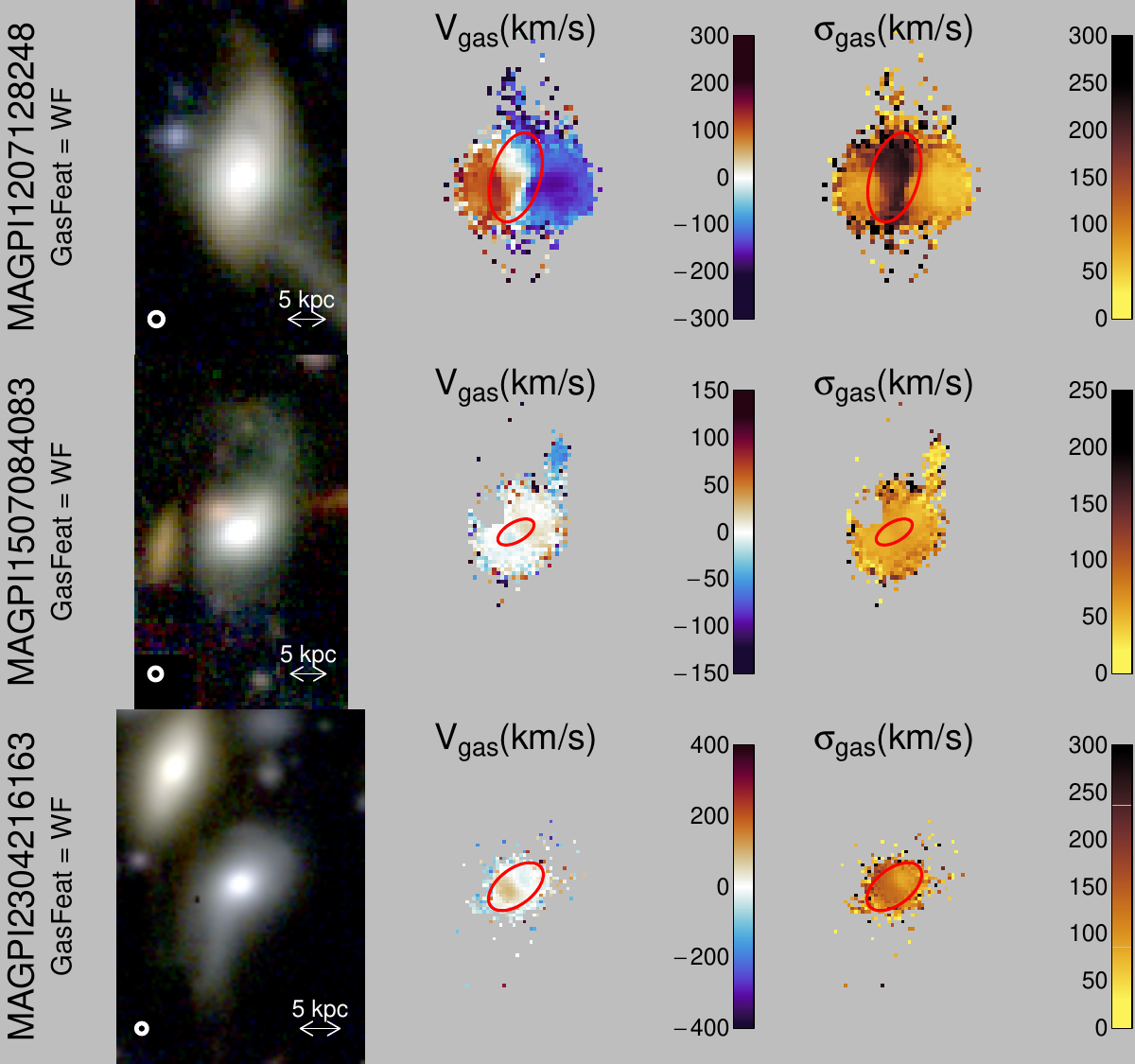}
\caption{Same as Figure \ref{fig:Stell_Feat}, but with gas kinematic feature(s). From top to bottom as labelled on the left: MAGPI1207128248 (Sp); MAGPI1507084083 (Merger); and MAGPI2304216163}\label{fig:Gas_Feat}
\end{figure}

\begin{figure*}
\centering
\includegraphics[width=18cm]{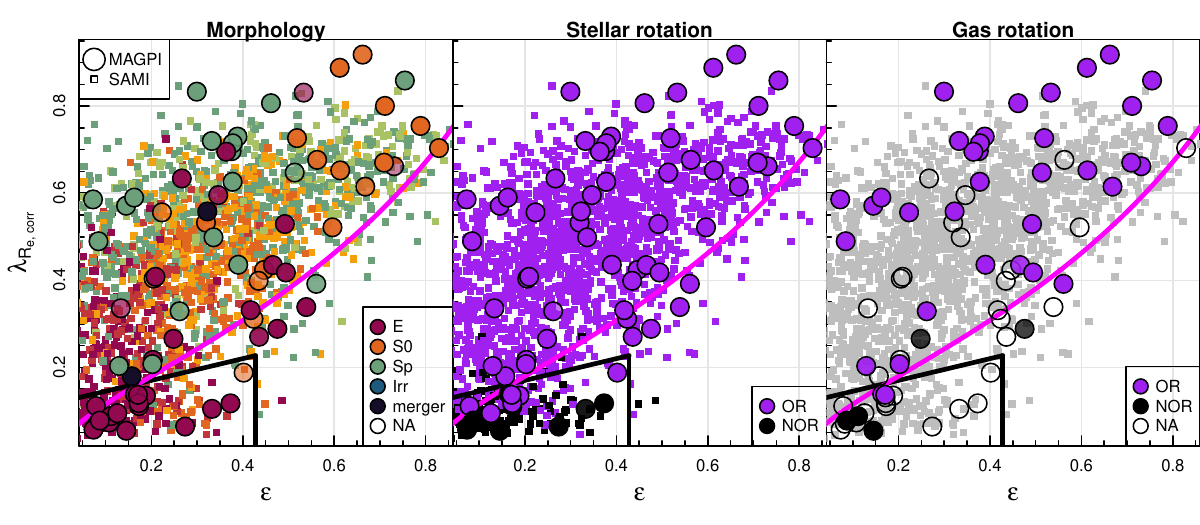}
\caption{Seeing-corrected spin-ellipticity diagram ($\lambda_{R_{\rm e, corr}}$ vs. $\epsilon$) for MAGPI (circles) galaxies as per \citet{Derkenne24} and SAMI (squares) galaxies as per \citet{vandeSande17a}. SAMI visual morphological and kinematic classifications are taken from \citet{Cortese16} and \citet{vandeSande21a}, respectively. Data are colour-coded by visual morphology (left), stellar rotation (middle) or gas rotation (right) as per the inset legend. Symbol transparency is inversely proportional to the respective posterior mode probability in each panel (left: $P_{\mu_{\rm Morph}}$, middle: $P_{\mu_{\rm StellOR}}$, and right: $P_{\mu_{\rm GasOR}}$). The black lines outline the division between fast and slow rotators suggested by \citet{vandeSande21a}. The magenta line shows the semi-empirical prediction for edge-on axisymmetric galaxies with anisotropy parameter $\beta= 0.70\epsilon_{\rm intr}$, where $\epsilon_{\rm intr}$ is the intrinsic ellipticity \citep[e.g.,][]{Cappellari07,Cappellari16}. The majority of galaxies that lie within the black lines are ellipticals with a high proportion of galaxies with NOR stars and gas. Hollow symbols are used when a classification is not available (NA).}\label{fig:spin-ell}
\end{figure*}

Figure \ref{fig:spin-ell} shows the distribution of MAGPI galaxies with $\log_{10}(M_\star/M_\odot)>10$ in the spin-ellipticity parameter space as per \citet{Derkenne24}. Most slow rotators (galaxies within the black lines) are classified as ellipticals and many show no obvious rotation (NOR) in either the gas (when available) or stars. There are some remarkable exceptions to these trends.
There are also 2 GasOR=NOR galaxies that are outside the slow rotator region in Figure \ref{fig:spin-ell}: MAGPI1534176099 and MAGPI2306197198. As indicated by the symbol transparency, neither of those classifications are deemed reliable, both having $P_{\mu_{GasOR}}<0.98$.

\begin{figure*}
\centering
\includegraphics[width=18cm]{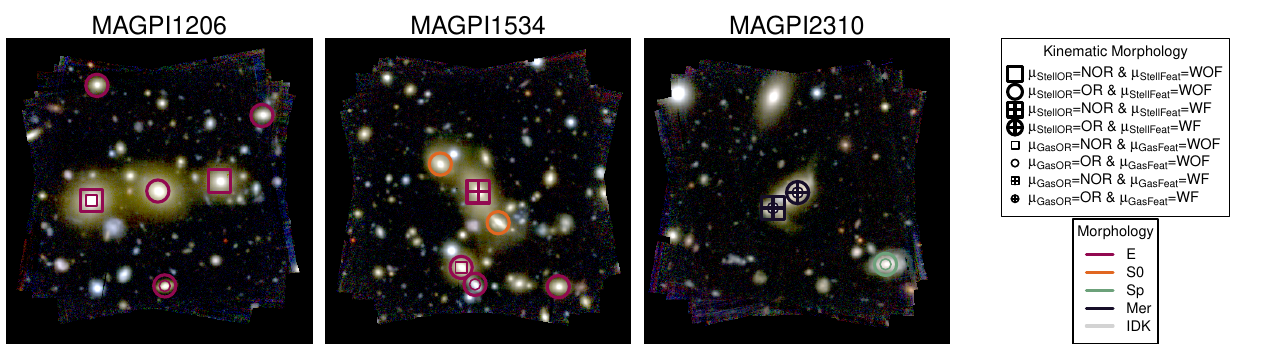}
\caption{MUSE synthetic $g$, $r$, $i$ images for three MAGPI fields included in this work with superimposed gas (small \& thin symbols, when present) and stellar (large \& thick) visual kinematic morphologies: OR (circles), NOR (squares), WOF (hollow), WF (cross). Colours correspond to the mode of the posterior morphological classification $\mu_{\rm Morph}$ (see legend on the right). Galaxies have a mix of visual and kinematic morphologies, with sometimes contrasting gas and stellar kinematic morphologies. A gallery of all fields is included in \ref{sec:appendix_fields_kinclass}.}\label{fig:kinclass_fields_subs}
\end{figure*}

Figures \ref{fig:kinclass_fields_subs} and \ref{sec:appendix_fields_kinclass} show that MAGPI galaxies have a mix of visual and kinematic morphologies, with sometimes contrasting gas and
stellar kinematic morphologies (also see Figure \ref{fig:Kin_Gallery}).
Figures \ref{fig:kinclass_fields_subs}, \ref{fig:morph-den} and \ref{sec:appendix_fields_kinclass} suggest that while the kinematic diversity of galaxies is already in place 3.5 billion years ago, the kinematic morphology-density relation is either too weak to be detected, yet to be established or possibly opposite to what is seen in local surveys. These options are considered in detail in Section \ref{sec:discussion}.

\section{Discussion}\label{sec:discussion}

Given the inherent subjectivity of visual classifications, we have chosen to collect multiple independent classifications on each galaxy and considered parameter. Each of the 637 galaxies included for visual classification has been classified by a minimum of 11 raters. We have found that there is important information in the spread of selected classifications and in the frequency of classifier hesitation to provide a classification that reflect the data quality and some properties of galaxies (i.e. how often classifiers leave a classification as `NS' or deliberately choose `IDK'). Studies using visual classifications involving few classifiers are thus subject to biases that are hard to quantify. Bayesian inference provides a statistically coherent way of quantifying our confidence in the true galaxy properties by combining a prior belief of classifier competence with the observed data via an explicit data-generating process. This analysis has also allowed us to quantify some of the conditions that make visual classifications more difficult.

%difficulties with morphological classifications
Previous local IFS surveys with comparable spatial resolution in physical units have available ancillary imaging with better spatial resolution and depth than the spectroscopy (e.g. SAMI \citealt{Cortese16} and MaNGA \citealt{VazquezMata22}). This is not the case for MAGPI, where both images and spectroscopy have the same spatial resolution, MAGPI data themselves usually represent the deepest and best images available for our targets. 
Difficulties associated with visual classification of small and/or faint galaxies has been raised in the literature \citep[see][for recent examples]{VazquezMata22,Tohill23,MedinaRosales24}.
The comparatively coarser spatial resolution of MAGPI images (in physical units) used to perform the visual morphological classifications has led to more uncertain morphological classifications in MAGPI compared to local surveys where comparatively higher quality images are available for visual classification. This has resulted in our inability to reliably differentiate between late and early spirals and difficulties in classifying faint and/or small objects, possibly leading to a bias towards classifying more amorphous earlier type morphologies. The difficulties associated with recognising small structural features in low resolution images is not limited to the current study and should be born in mind for future studies using visual morphological classifications.

Despite these difficulties, for our bright sample ($r<20$ mag), we find that earlier types have higher S\'ersic indices than later types (Figure \ref{fig:Morph-n}) as expected. This is evidence that despite their higher uncertainties, the morphological classifications are informative, especially when considered as a sample. %We may in the future use morphologies derived in this work to help fine-tune \citep[e.g.][]{Omori23} pre-trained machine learning tools such as {\sc ZooBot} \citep{Walmsley22a,Walmsley22b} to improve existing and derive reliable morphological classifications for the remaining unclassified MAGPI sample. 

%Difficulties with kinematic classifications
The distributions in posterior mode probabilities shown in Figure \ref{fig:Mode_hist} and sample sizes listed in Table \ref{table:families}, along with the dependence of the difficulty on brightness and size (Figure \ref{fig:difficulties}), suggest that at comparable data quality and spatial resolution, bulk rotation is easier to identify than visual morphology. We find that kinematic features (StellFeat and GasFeat), which often represent small scale kinematic anomalies are easier to visually identify than the absence of rotation (i.e., the mean difficulty of StellOR=NOR and GasOR=NOR is higher than the mean difficulty for StellFeat=WF and GasFeat=WF, respectively), although $N_{\rm pix}>100$ appears to be a lower threshold for identifying StellFeat=WF and GasFeat=WF. This has limited our ability to quantitatively compare our kinematic morphologies of galaxies with samples at different depths and/or spatial resolution \citep[e.g.][]{Krajnovic11,vandeSande18}. Qualitatively, we were able to identify reliable ($P_{\mu}>0.98$) representatives of nearly all classes within MAGPI (exceptions being irregular and late type spiral morphologies, see examples in Figures \ref{fig:Kin_Gallery} and \ref{fig:Morph_Gallery}).

%SAMI comparison.
The spatial resolution of MAGPI and SAMI kinematic maps are comparable (SAMI median is $\sim1.9$ kpc/FWHM, while the MAGPI bright sample has median $\sim2.7$ kpc/FWHM) and we have closely followed the methodologies used in SAMI for stellar kinematic visual classifications (i.e. StellOR and StellFeat).
We thus compile a methodologically comparable dataset of local ($z\sim 0$) galaxies with $\log_{10}(M_\star/M_\odot)>9.5$ dex from the SAMI Galaxy Survey \citep{Croom12,Croom21}. We use visual morphologies from \citet{Cortese16}, spin values from \citet{vandeSande17b} and visual stellar kinematic morphologies from \citet{vandeSande21a}. We combine this with GAMA group masses and environment metrics based on \citet{Robotham11} (assuming $A=10$ for consistency). Figure \ref{fig:SAMI_selection_comp} compares the stellar mass, group mass, effective radius and S\'ersic index distributions of SAMI and MAGPI galaxies. We note that given the differences in stellar mass distributions, the properties of galaxies in the SAMI sample are not representative of those of the descendants of MAGPI galaxies, thus a direct evolutionary link cannot be inferred between the two samples.

Without referring to evolution, we note that there are distinct similarities between the distributions of SAMI and MAGPI galaxies in the spin-ellipticity parameter space (see Figure \ref{fig:spin-ell}), with elliptical galaxies dominating the slow-rotator area (i.e. black lines) and later-types (S0s and Sps) occupying the fast-rotator areas. This shows that locally observed trends with visual morphologies are broadly established 3.5 billions years ago.
Indeed, \citet{vanHoudt21} have shown that a comparable trend with S\'ersic index is noticeable in $v/\sigma-\epsilon$ space at even higher redshift ($z\sim 0.8$) in the LEGA-C survey using dynamical Jeans models of slit spectra.

Similarly, as seen in SAMI, MAGPI galaxies with/without obvious stellar rotation tend to occupy or lie near the fast-/slow-rotator area in the middle panel of Figure \ref{fig:spin-ell}. 
A study of the spin-evolution of galaxies through comparing carefully selected sub-samples of MAGPI descendent-like MaNGA galaxies is presented in \citet{Derkenne24}.

\begin{figure*}
\centering
\includegraphics[width=18cm]{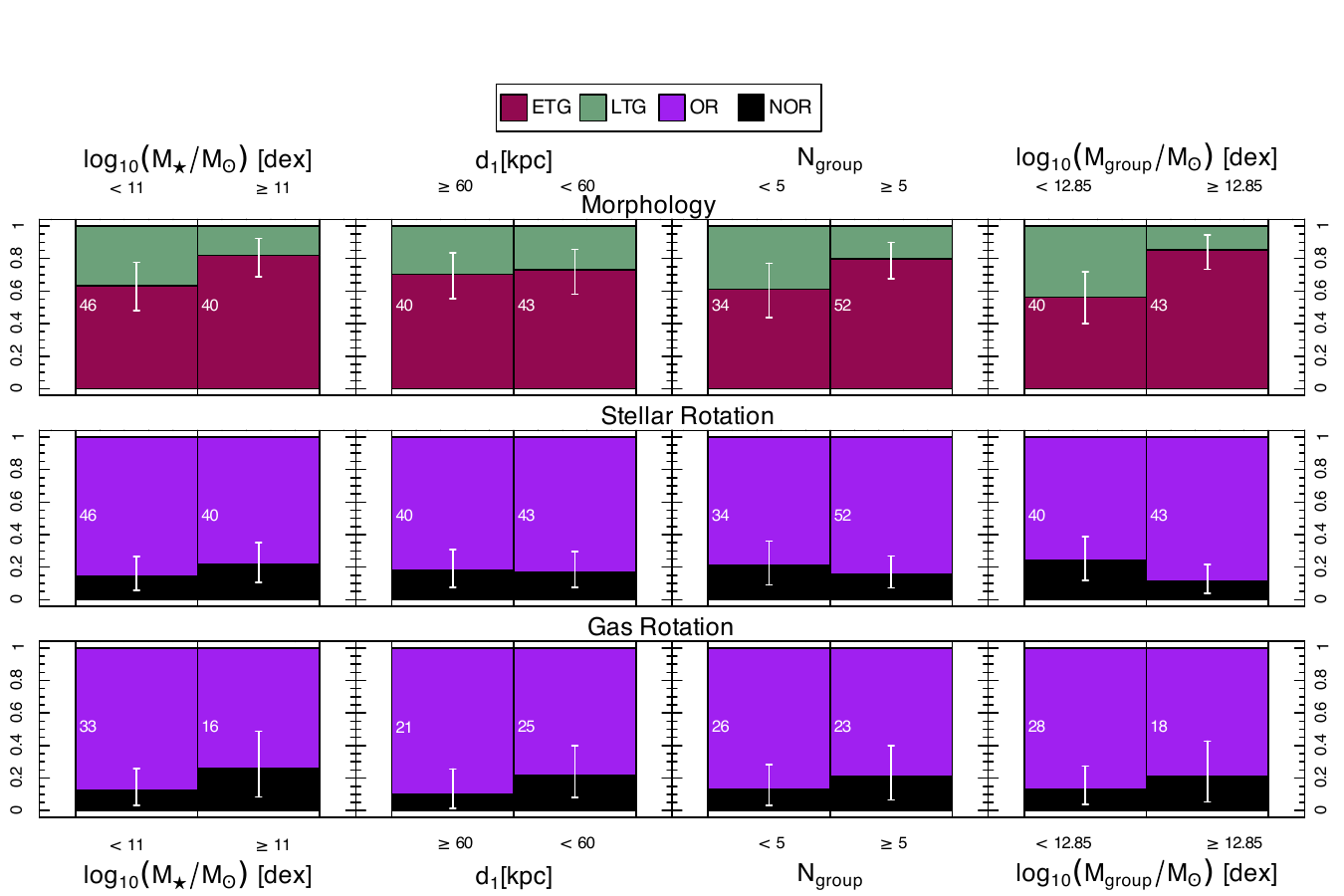}
\caption{Proportion of early- (red) vs. late- (green) type galaxies (top), stellar (middle) and ionised gas (bottom) with (purple) vs. without obvious (black) rotation (see legend at the top) for the bright sample. Plots show stellar mass (left) and environmental bins for three separate proxies: 1- divided according to the distance to their nearest neighbour ($d_1$, middle left): densest neighborhood ($d_1<60$ kpc, within the range of commonly used thresholds, e.g. \citealt{Robotham14}) and lowest density neighborhood ($d_1\ge60$ kpc); 2- divided according to the number of galaxies in their group (middle right): isolated and/or small group ($N_{\rm group}<5$) and larger groups ($N_{\rm group}\ge5$); and 3- group dynamical mass proxy divided at the median group mass of ($\log M_{\rm group}/M_\odot=12.85$, right). The number of galaxies in each environment bin is shown in white and Bayesian 95 percent credible intervals are shown as white errorbars (see Table \ref{table:resampling}). %Fractions are currently indistinguishable across environment, i.e. there is no detectable environmental trend.
}\label{fig:morph-den}
\end{figure*}

\begin{figure}
\centering
\includegraphics[width=8.5cm]{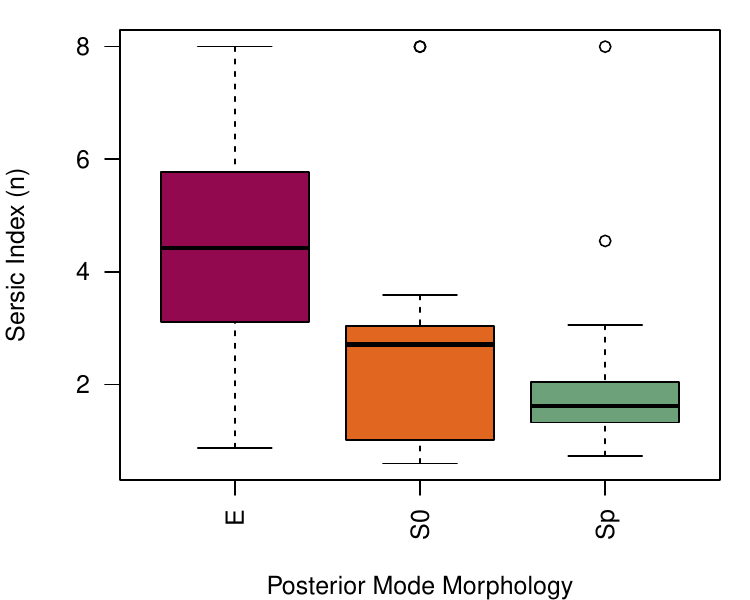}
\caption{Box and whiskers plot showing the quartiles of the distributions of S\'ersic indices (n) for bright ($r<20$ mag) E, S0 and Sp (including both eSp and lSp categories) MAGPI galaxies. As expected, the distribution for elliptical galaxies is skewed towards higher S\'ersic indices, while that of Spiral galaxies is skewed towards lower values. Colours are consistent with those of the left-panel of Figure \ref{fig:spin-ell}.}\label{fig:Morph-n}
\end{figure}

\begin{figure}
\centering
\includegraphics[width=8.5cm]{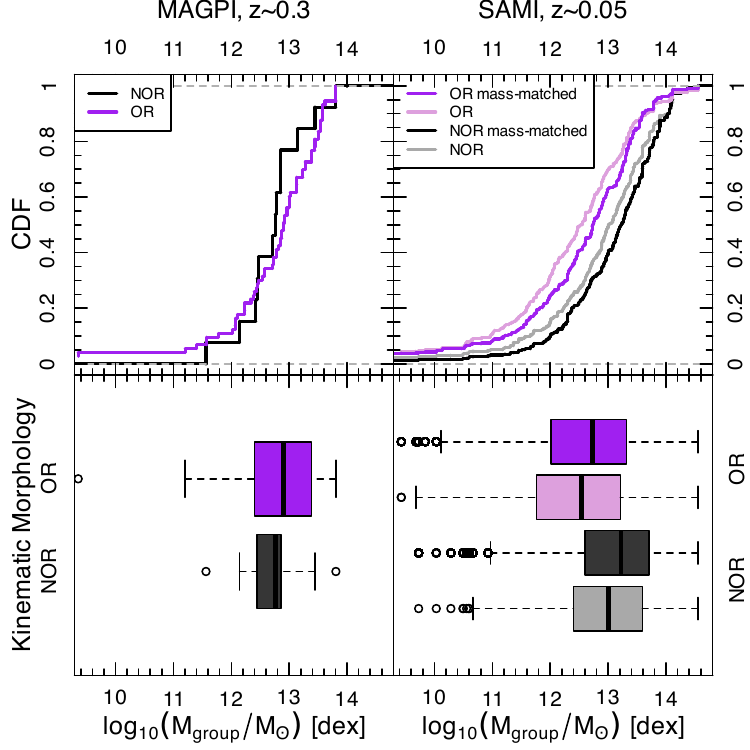}
\caption{Group mass cumulative density function (top) and whiskers plot (bottom) for the MAGPI bright sample (left) and SAMI GAMA sample (i.e. excluding cluster galaxies, right). For MAGPI, we use only galaxies with reliable $\mu_{\rm StellOR}$ (i.e. $P_{\mu_{\rm StellOR}}>0.98$), while for SAMI we use galaxies with $M_\star>10^{9.5}M_\odot$ and available kinematic morphology from \citet{vandeSande21a}. We show the original SAMI sample in lilac (OR) and light grey (NOR) as well as a sample mass-matched to the MAGPI sample in purple (OR) and black (NOR). The relative $M_{\rm group}$ distributions for NOR (black) and OR (purple) galaxies suggest that galaxies without obvious stellar rotation prefer lower group masses than galaxies with obvious rotation in intermediate redshift MAGPI galaxies, while the opposite is true for low redshift SAMI galaxies.}\label{fig:ORNORenvCDF}
\end{figure}

In what follows, we implicitly assume that an ``evolved'' galaxy is a massive, old, quenched, non-rotating galaxy. We note that  evolution to this state needs not be linear (apart for mass) and the order in which those attributes appear will likely depend on the individual galaxy's merger and star formation history.

While earlier studies of the kinematics and morphologies of galaxies at cosmic noon ($z\sim1-3$) found a predominance of star forming clumpy and discy rotating systems \citep[e.g.][but see \citealt{Rodrigues17}]{Wuyts11,Wisnioski15,HuertasCompany16,Stott16}, most of dynamical studies necessarily focused on ionised gas as the dynamical tracer. In the local universe, ionised gas and active star formation are predominantly found in late-type systems with discy dynamics. We will briefly assume that a similar link between star formation, morphology and dynamics holds in the early Universe. Recent studies from the James Webb Space Telescope images suggest spheroidal morphologies are the dominant type for masses $\log_{10}M_\star/M_\odot > 11$ galaxies in the early Universe \citep[up to $z<8$, see e.g.][]{Lee24}, and a quiescent system was found at $z\sim11$ \citep{Glazebrook24}.
If seemingly ``evolved'' spheroidal and/or quiescent systems are in place early (i.e. at $z>>0.3$) in the history of the Universe, and broadly assuming visual structure mirrors their dynamics as seen in local galaxies, it follows that a commensurate population of dynamically evolved potentially non-rotating galaxies is found in our $z\sim0.3$ sample.
Indeed, \citet{Lagos18b} show that the median spin of slow rotators at redshift $z\sim0.3$ is only marginally higher than that at $z\sim0$. We note that when comparing the distribution of $\lambda_{R_e}$ as a function of redshift in three separate cosmological simulations, \citet{Foster21} find that different cosmological simulations obtain distinct predictions at $z\sim0.3$.

The following discussion focuses on stellar kinematics because there is comparable work in the literature. While a population of non-rotators is present at $z\sim0.3$ \citep[also see][]{Derkenne24}, the fraction of NOR galaxies in environmental bins parametrised by $N_{\rm group}$ and $M_{\rm group}$ (in stars) is marginally higher in lower density environment bins (see Figures \ref{fig:morph-den} and \ref{fig:ORNORenvCDF}). This is in agreement with \citet{MunozLopez24}, who did not find evidence that galaxies with higher spins prefer less dense environments at intermediate redshift, but in contrast to that found in local surveys \citep[e.g.][]{Cappellari11b,DEugenio13,Fogarty14,Brough17,Greene17}. We note that the kinematic morphology-density relation was first found at $z\sim0$ in the ATLAS$^{\rm3D}$ survey \citep{Cappellari11b} in a sample roughly 2.5 times the size of that presented here, and with better spatial resolution than that of MAGPI. \citet{Fogarty14} did detect the kinematic morphology-density relation in Abell 85 using SAMI data, but not in the Abell 168 and 2399 clusters. We next discuss the statistical significance of this result and ascertain whether the differences in resolution and sample size could explain the absence of the stellar kinematic morphology-density relation in MAGPI or whether this result is robust.

%#[1] pvalue_envtrend=0.17455 #consistent with no trend or reverse trend with 82 galaxies (i.e. about 60% the final sample.)
%# pvalue_envextr=<0.00001 #consistent with as extreme or more extreme a reverse trend as MAGPI
%#[1] pvalue_envtrend=0.09887 #consistent with no trend or reverse trend with 82 galaxies if also excluding the SAMI cluster fields.
%# pvalue_envextr=<0.00001 #consistent with as extreme or more extreme a reverse trend as MAGPI if also excluding the SAMI cluster fields
%#[1] pvalue_envtrend=0.10932 #consistent with as extreme or more extreme a reverse trend as MAGPI with full MAGPI sample of 58 fields (which will include archive cluster fields)
%# pvalue_envextr=0.00002

The stellar kinematic morphology-density relation has been shown to be of secondary importance, after the mass \citep[e.g.][]{Greene17,vandeSande21b,Vaughan24} and age \citep{Croom24} in local samples. Nevertheless, when one does not account for stellar mass or age (as done here), the environmental trend is clear locally, e.g. using the SAMI dataset \citep[][and Figure \ref{fig:ORNORenvCDF} of the current work]{Brough17}. 

However, since there is in fact a lower fraction of NOR in high mass groups, the proportions of stellar NOR galaxies with group mass is reversed in MAGPI compared to SAMI. We want to ascertain our confidence in this result in a way that accounts both for the data and sample size. For each environment tracer ($d_1$, $N_{\rm group}$ and $M_{\rm group}$) and environmental bin (low- and high-density), we calculate the posterior distribution of the prevalence of galaxies with StellOR=NOR, based on a non-informative ${\rm Beta}(1,1)$ prior distribution (i.e. we assume a completely non-informative prior belief that the prevalence of NOR galaxies in each bin could be anywhere between 0 and 1) and the Bayesian model for their classifications. This accounts for sampling because as the sample size gets larger, the posterior shifts away from the prior and puts more weight on the observed data. From this, we calculate the posterior probability that the prevalence of galaxies in the low-density bin ($f_{\rm NOR,LD}$) with StellOR=NOR is higher than that in the high-density bin ($f_{\rm NOR,HD}$). When dividing galaxies by group mass, the probability of there being either no or an inverse trend in MAGPI ($p_{\Delta f_{\rm NOR}>0}=1-p_{\Delta f_{\rm NOR}<0}$ where $\Delta f_{\rm NOR} = f_{\rm NOR,LD}-f_{\rm NOR,HD}$) in Table \ref{table:resampling}) is 90.9 and 94.4 percent for the whole and bright samples, respectively. When galaxies are split by the number of group members, the probabilities are 98.5 and 72 percent for the whole and bright samples, respectively. When galaxies are split by $d_1$ and all galaxies are included, the posterior probability is higher still. This suggests there is some evidence that galaxies with StellOR=NOR may be more common or present in similar proportions in the low-density environment bins in MAGPI when galaxies are divided by group mass or number of group members.

To test the significance of the seeming discrepancy between the kinematic morphology-density relation seen at low redshift and our MAGPI results, we obtain 100,000 samples of 83 SAMI galaxies with agreed upon stellar kinematic morphology classifications taken from \citet{vandeSande21a}. Each sample is matched to the mass distribution of MAGPI. For each sample, we record $f_{\rm NOR,HD}$ and $f_{\rm NOR,LD}$ with a division between high and low density at $\log(M_{\rm group}/M_\odot)=12.85$. This implicitly assumes no evolution between SAMI and MAGPI. Since SAMI did not use a Bayesian framework as used here to obtain posterior probabilities for each classification, we assume that the posterior probability for all SAMI kinematic morphology classifications is uniformly 1. We note that the different choices of methodologies limit our ability to perform a fair comparison and this must be taken as a caveat (also relevant to Figure \ref{fig:ORNORenvCDF}). We then take 1,000,000 samples of the posterior distributions and find a median and 95 percent credible interval for $\Delta f_{\rm NOR} = -0.24(-0.46 - -0.01)$.
The posterior of the probability $p_{\Delta f_{\rm NOR}>0}$ that $\Delta f_{\rm NOR}$ is positive as seen in MAGPI is strongly skewed at 0 with median $0.019$ and 95 percentile $0.33$, suggesting that an equal or higher fraction of NOR in low group mass bins would typically randomly occur $\sim1.9$ percent of the time in SAMI.

We also look for how often we find a difference in the NOR fraction that is opposite and as or more extreme as that measured in MAGPI. Again, we find that $p_{\Delta f_{\rm NOR}>0.129}$ is strongly peaked at 0, with a median of $=0.00070$ and 95 percentile of $0.061$, suggesting strong tension between MAGPI and SAMI in the proportions of NOR in low vs. high mass groups. These results qualitatively hold if we exclude cluster galaxies from the SAMI sample (median $p_{\Delta f_{\rm NOR}>0}=0.0082$, %mistakenly quoted the means rather than median in original submission. Also boosted stats with Nmc=1,000,000 (factor of 10 larger)
and $p_{\Delta f_{\rm NOR}>0.129}=0.00016$); when using a sample size comparable to that expected for the complete MAGPI survey (i.e., sampling 136 galaxies, median $p_{\Delta f_{\rm NOR}>0}=0.0040$ and $p_{\Delta f_{\rm NOR}>0.129}=0.000023$); or when matching both the MAGPI stellar mass and S\'ersic index distributions simultaneously (median $p_{\Delta f_{\rm NOR}>0}=0.025$ and $p_{\Delta f_{\rm NOR}>0.129}=0.00092$). In other words, if the kinematic morphology-density relation was similar in both SAMI and MAGPI, we would very likely have detected it. Furthermore, it is highly unlikely that a SAMI-like kinematic morphology-density relation at $z\sim0.3$ would lead to the result found in MAGPI.

Simulations also are equivocal about the role of environment in producing slow rotators. Using over 10,000 simulated galaxies in the Horizon-AGN simulation, and echoing the results found in this work, \citet{Choi18} found similar proportions of slow rotators across all environmental bins, with slow rotation typically resulting from non-merger tidal perturbations in lower density environments. They do however see a weak trend with satellites vs. centrals. While some residual environmental effects were detected after accounting for mass, \citet{Lagos18b} find that $\sim30$ percent of slow rotators in over 16,000 EAGLE and HYDRANGEA galaxies have not experienced a recent merger, but their slow rotation can be instead attributed to their residing in lower spin halos than slow rotators formed through major mergers. 
The multiple formation pathways in simulations indicate that environmental effects need not be the only driver of slow rotation. Comparison of the specific angular momentum of the dark matter halo to that of the stars and HI gas in observed and simulated galaxies confirm that the dynamics of dark matter halo impact those of baryons \citep[e.g.,][also see \citealt{Fall23}]{El-Badry18,Romeo23}.

Relating the 3D shape of slow rotators to their environments, \citet{Lagos22} find that intrinsically flat satellite slow rotators are more common in low mass halos, while prolate slow rotators are more common in high mass halos. It may be that the MAGPI selection has favoured flat slow rotators over prolate or round ones. Distinguishing intrinsically prolate, spherical and flat galaxies is challenging observationally \citep[e.g.][]{MendezAbreu16,Bassett19,Yong24} and beyond the scope of this work. 

The theoretical results outlined above do not rule out the possibility that the preponderance of slow rotation in satellite galaxies may be distinct from that in centrals. The MAGPI selection of fields centered on massive centrals may have skewed the sample towards more massive centrals and less massive satellites than in the general population. This in turn could prevent a fair comparison should slow rotation strongly depend on central vs. satellite status. Indeed, \citet{Croom24} found that the spin of centrals and satellites had distinct dependencies on environment. We state this as a possible caveat of our analysis. In \citet{Foster25a}, we separately account for satellite and central status and find that the impact of environment on centrals and satellites can differ depending on the dynamical parameter studied.

We note that while $d_1$ is helpful in identifying potentially interacting galaxies, it is not otherwise a recommended tracer of environment density. Instead, the distance to the $N^{\rm th}$ nearest neighbour (which is not used here as it is not yet available for a large proportion of our galaxies), where $N\sim5$ is more commonly used \citep[e.g.][]{Muldrew12,Brough13,deVos24}. This may further explain why results for $d_1$ differ from other measures of environment densities considered herein (i.e. $N_{\rm group}$ and $M_{\rm group}$).

\begin{table}
\caption{Inferred percentage (\%$_{\rm NOR}$) and respective 95 percent credible interval (CrI) of galaxies without obvious stellar rotation (i.e. StellOR=NOR) in high- (column 3) and low-density environments (column 5) in MAGPI. For each environment parameter and environmental bin the number $N$ of galaxies is given in columns 2 and 4. The test is performed on both the whole and bright ($r<20$ mag) samples (column 1). The $\Pr({\rm High} > {\rm Low})$ value given in column 6 is the posterior probability that there is a higher prevalence of non-rotating galaxies in the denser environment, as seen in the local universe.}
\centering
\resizebox{\columnwidth}{!}{\begin{tabular}{| l | c c | c c | c|}
\hline
Sample&\multicolumn{2}{|c|}{High density} & \multicolumn{2}{|c|}{Low density} &\\
&$N$ & $f_{\rm NOR}$ (CrI) & $N$ & $f_{\rm NOR}$ (CrI) & $p_{\Delta f_{\rm NOR}<0}$\\
(1) & (2) & (3) & (4) & (5) &(6)\\
\hline\hline
&\multicolumn{2}{|c|}{$d_1<60$} & \multicolumn{2}{|c|}{$d_1\ge60$}&\\
\hline
All&62&0.318 (0.208 – 0.438)&73&0.287 (0.183 – 0.403)&0.65\\
Bright&43&0.170 (0.074 – 0.295)&40&0.178 (0.076 – 0.310)&0.46\\
\hline\hline
&\multicolumn{2}{|c|}{$N_{\rm group}\ge5$} & \multicolumn{2}{|c|}{$N_{\rm group}<5$}&\\
\hline
All&84&0.269 (0.177 – 0.372)&72& 0.436 (0.318 – 0.555)&0.015\\
Bright&52&0.159 (0.074 – 0.268)&34&0.210 (0.091 – 0.360)&0.28\\
\hline\hline
&\multicolumn{2}{|c|}{$\log_{10}M_{\rm group}\ge12.85$} & \multicolumn{2}{|c|}{$\log_{10}M_{\rm group}<12.85$}&\\
\hline
All&69&0.249 (0.152 – 0.360)&66&0.355 (0.240 – 0.479)&0.091\\
Bright&43&0.112 (0.038 – 0.218)&40&0.241 (0.119 – 0.388)&0.056\\
\hline
\end{tabular}}
\label{table:resampling}
\end{table}

If we take the absence of a stellar kinematic morphology-relation in MAGPI at face value, our findings could indicate a scenario in which many non or slowly rotating MAGPI galaxies have yet to join or their surrounds have yet to grow into high density environments, where such galaxies tend to preferentially be at $z\sim0$. In other words, we speculate that galaxies may grow their mass first and spin down sometime before $z\sim0.3$, they subsequently either join dense environments or their environment becomes denser over time. Similarly, \citet{MunozLopez24} found that spin is not correlated with environment in an independent sample of fast rotators (they did not find any slow rotators) at intermediate redshift ($0.1 \le z \le 0.8$). Their sample did not however include the highest density environments.

%From Claudia: It may be worth adding here the typical assembly time of galaxy clusters to emphasize the fact that the largest structures are the latest to form (e.g. Amoura et al. 2021) -> in that paper they show that galaxy clusters (1e15Msun) reach 50% of their z=0 mass only at z~0.35. If you go to massive groups (1e13Msun) then 50% is reached at z~0.6. So basically it makes sense that by z=0.3 we are yet to see a lot of the mos massive clusters forming.

Under the $\Lambda$CDM paradigm, the largest structures are the latest to form. \citet[][see their figure 3]{Amoura21} show galaxy clusters ($M>10^{15}M_\odot$) reach 50 percent of their present day ($z=0$) mass around $z\sim0.35$. Massive galaxy groups ($M>10^{13}M_\odot$) reach 50 percent of their present day mass around $z\sim0.6$. Thus, at intermediate redshift $z\sim0.3$, a large proportion of the most massive clusters are yet to form.

Environment was already shown to be a second-order effect on determining the rotation state of galaxies, with age being the strongest determinant for stellar spin in local galaxies \citep{Croom24}. In the scenario painted above, if galaxies spin down as they age and grow before their environment is fully assembled, the emerging environmental trend seen in local surveys could be weaker than the mass trend and will likely get stronger as the Universe continues to evolve and structures continue to assemble. 

This scenario is reminiscent of the pre-processing (usually referring to star formation quenching) of galaxies in groups or cosmic filaments prior to joining clusters \citep[e.g.][]{Fujita04,Sarron19,Sengupta22}. We suggest that galaxies without obvious rotation may have experienced ``dynamical pre-processing'' in lower mass halos before progressively being added to higher mass halos after $z\sim0.3$.

\section{Conclusions}\label{sec:conclusions}
In this work, we present the results of morphological and dynamical visual classifications for 637 spatially resolved galaxies within 35 MAGPI fields. 
From this initial sample, we select sub-samples of bright (r < 20 mag) high-confidence (> 0.98) morphological (86 galaxies) and stellar kinematic (82 galaxies) classifications at redshifts ($0.2 \le z \le 0.4$).
Our aim is to identify and quantify the dynamical state of galaxies at intermediate redshifts and look for signs of evolution through comparing with the comparable IFS SAMI Galaxy Survey at $z\sim0$.

Our conclusions are as follows:
\begin{enumerate}
\item Despite the difficulty of detecting small scale structures in moderate spatial resolution data, we find examples of a range of morphological types present at intermediate redshift corresponding to $\sim3.5$ Gyr lookback time.
\item Similarly, galaxies with and without obvious rotation are already in place at intermediate redshift. The fractions of galaxies with and without obvious rotation mirror those of fast and slow rotators, respectively. 
\item We do not find a positive trend in the fraction of non-rotating galaxies with group mass (i.e. the stellar kinematic morphology-density relation) within the studied sample in contrast to local studies.
\end{enumerate}

An important caveat is that the MAGPI selection favours lower mass galaxies being satellites of massive centrals, which may have an impact on the dynamical flavours of galaxies included in the sample, especially at lower masses.

If taken at face value, we suggest that the absence of an environmental trend with stellar rotation at $z\sim0.3$ (compared to local studies) is consistent with results indicating at most a secondary role for environment. Indeed, environment has been shown to be secondary after mass \citep[e.g.][]{Greene17,Brough17,vandeSande21b,Vaughan24} and potentially stellar age \citep{Croom24} for low redshift IFS surveys. We further expect that environmental trends become more prominent as large scale environments continue to build up as the Universe evolves. In other words, the presence and similar proportions of non-rotating galaxies in both environment bins suggest that galaxies with no obvious stellar rotation were dynamically pre-processed at an earlier epoch. The kinematic morphology-density relation has not yet emerged by $z\sim0.3$, suggesting the non-rotating galaxies already present have yet to join denser environments. This will be contrasted with theoretical expectations using mock observations of cosmological simulations.

\begin{acknowledgement}
%Individuals
LMV acknowledges support by the German Academic Scholarship Foundation (Studienstiftung des deutschen Volkes) and the Marianne-Plehn-Program of the Elite Network of Bavaria.

%ESO
Based on observations collected at the European Organisation for Astronomical Research in the Southern Hemisphere under ESO program 1104.B-0536. We wish to thank the ESO staff, and in particular the staff at Paranal Observatory, for carrying out the MAGPI observations. 

%MAGPI
MAGPI targets were selected from GAMA. GAMA is a joint European-Australasian project based around a spectroscopic campaign using the Anglo-Australian Telescope. GAMA is funded by the STFC (UK), the ARC (Australia), the AAO, and the participating institutions. GAMA photometry is based on observations made with ESO Telescopes at the La Silla Paranal Observatory under programme ID 179.A-2004, ID 177.A-3016.

%CMasher for good-lookin' colour scales.
This work makes use of colour scales chosen from \citet{CMASHER2020}. 
\end{acknowledgement}

\paragraph{Funding Statement}

%Individual grants
CF is the recipient of an Australian Research Council Future Fellowship (project number FT210100168) funded by the Australian Government.
CL, JTM and CF are the recipients of and KH acknowledges funding from the Australian Research Council (ARC) Discovery Project DP210101945.
AFM was supported by RYC2021-031099-I and PID2021-123313NA-I00 of MICIN / AEI / 10.13039 / 501100011033 / FEDER,UE, NextGenerationEU/PRT
FDE acknowledges support by the Science and Technology Facilities Council (STFC), by the ERC through Advanced Grant 695671 ``QUENCH'', and by the UKRI Frontier Research grant RISEandFALL.
%ASTRO3D
Part of this research was conducted by the Australian Research Council Centre of Excellence for All Sky Astrophysics in 3 Dimensions (ASTRO 3D), through project number CE170100013.

\paragraph{Competing Interests}

None.

\paragraph{Data Availability Statement}

The MAGPI raw data (and a basic data reduction) are available through the \href{http://archive.eso.org/cms.html}{ESO Science Archive Facility}.

%A statement about how to access data, code and other materials allowing users to understand, verify and replicate findings --- e.g. Replication data and code can be found in Harvard Dataverse: \url{https://doi.org/link}.

%\endnote in some journals will behave like \footnote; and \printendnotes will not output anything. 
%\printendnotes

\bibliography{biblio}

\appendix

\section{Extended Dawid--Skene model}\label{sec:appendix_dsmodel}

We suppose that the given property of galaxy $j$ has true category $z_j \in \{1, \ldots, K \}$, where the prevalence of categories in the population is characterised by the vector $\boldsymbol{\pi} = (\pi_1, \ldots, \pi_K)$ such that $z_j \sim \mathrm{Categorical}(\boldsymbol{\pi})$.

Let $d_j \in (0, 1)$ denote a measure of ``difficulty" in visually classifying galaxy $j$, and $c_i \in \mathbb{R}$ a measure of relative ``confidence" for classifier $i$ to provide a classification. We model the probability that classifier $i$'s response $Y_{ij}$ when shown galaxy $j$ is NS or IDK as

$$\Pr(Y_{ij} \in \{\mathrm{NS}, \mathrm{IDK}\}) = \frac{d_j \exp(-c_i)}{1 - d_j(1 - \exp(-c_i))},$$

such that $d_j$ can be interpreted as the probability that the ``average" classifier (with $c_i = 0$) would not provide a classification, and $c_i$ is the logarithm of the odds ratio comparing the probability that classifier $i$ provides a classification for a given galaxy to that of the average classifier.

When providing a classification, classifier $i$'s response when shown a galaxy with true parameter $z$ is characterised by the ``response tendency" vector $\boldsymbol{\theta}_{iz} = (\theta_{iz1}, \ldots, \theta_{izK}) \in \Delta^K$. Specifically,

$$\Pr(Y_{ij} = k \mid Y_{ij} \notin \{ \mathrm{NS}, \mathrm{IDK} \}) = \delta(\theta_{iz_jk}, d_j)$$

where $\delta: [0,1]^2 \mapsto [0,1]$ is a function which describes how these tendencies approach a uniform distribution as difficulty increases: 

$$\delta(\theta, d) = \frac{\theta(1 - d) + d\alpha}{1 - d(1 - K\alpha)}$$

such that $\delta(\theta, 0) = \theta$ (so $\boldsymbol{\theta}_{iz}$ can be interpreted as the response tendencies for classifier $i$ when shown a galaxy of type $z$ with hypothetical zero difficulty), $\delta(\theta, 1) = 1/K$ and $\alpha > 0$ controls how quickly the diffusion towards uniformity occurs as difficulty increases.

For Bayesian inference, we used weakly informative priors. For $\boldsymbol{\pi}$ we used a symmetric Dirichlet prior with all parameters equal to 1. For each $\boldsymbol{\theta}_{iz}$ we used a Dirichlet prior with parameter equal to $7(K-1)/6$ on the correct classification and $1/2$ on each incorrect classification. Each difficulty $d_j$ was modelled as being drawn from a Beta distribution, with a $\mathrm{Uniform}(0, 1)$ hyperprior on the mean parameter and a $\mathrm{Pareto}(0.1, 1.5)$ hyperprior on the `sample size' parameter \citep[][Chapter 5]{Gelman13}. Each relative confidence $c_i$ was modelled as being drawn from a normal distribution, with zero mean and a half-normal hyperprior with scale $3/(\Phi^{-1}(0.975)\Phi^{-1}(0.995)) = 0.594$ on its standard deviation. For the logarithm of $\alpha$, we used a half-normal prior with scale $2/\Phi^{-1}(0.995) = 0.776$.

For each parameter, we ran 8 Markov chains each for at least 13,000 steps following 2,000 steps of adaptive burn-in. We performed diagnostic checks to ensure that the algorithm successfully explored the posterior and that posterior quantities were well-estimated. For each classifier, for each galaxy, and overall, we compared the posterior predictive distribution of responses to the observed data in order to confirm the adequacy of the model \citep[][Chapter 6]{Gelman13}. For each galaxy $j$, we extracted the posterior mean of the difficulty parameter, $D = \mathbb{E}(d_j \mid \boldsymbol{Y})$, and the posterior probabilities that the true category corresponded to each option $k$, $P_k = \Pr(z_j = k \mid \boldsymbol{Y})$. The posterior mode was the category with the largest posterior probability, $\mu = \argmax_k P_k$.

\section{Failed classifications}\label{sec:appendix_threshold}

\begin{figure}
\includegraphics[width=8.5cm]{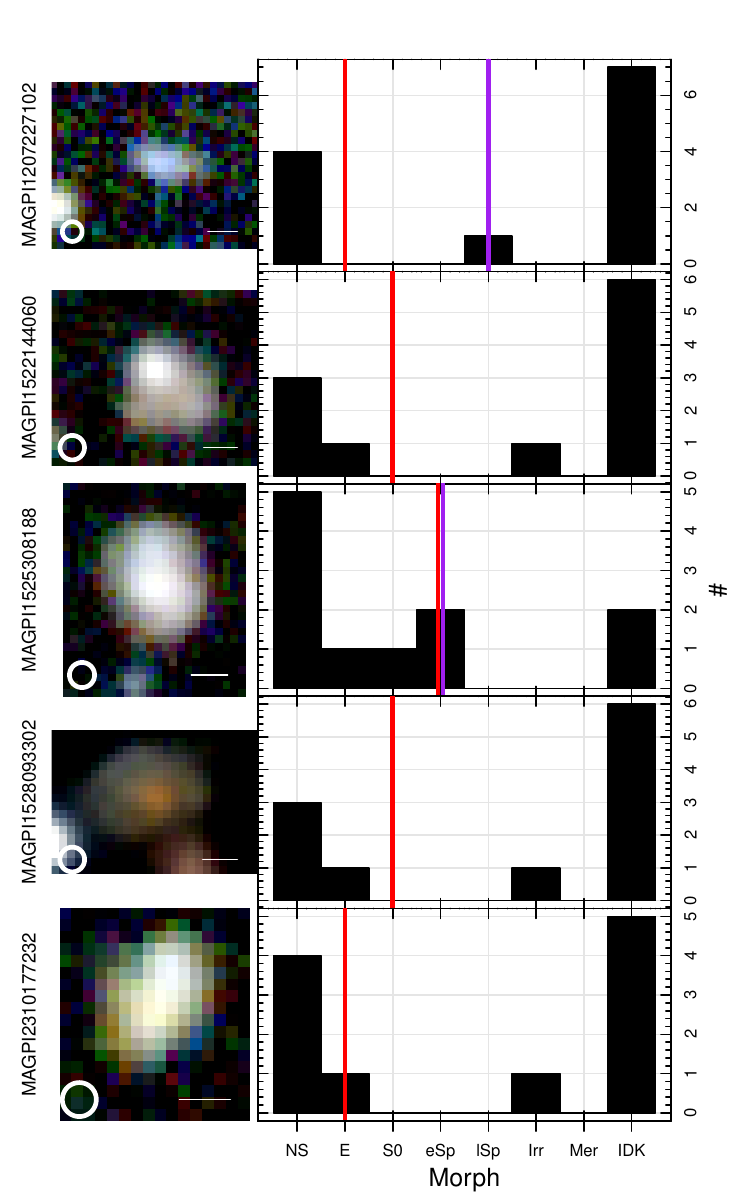}
\caption{Example MAGPI galaxies (as labelled on the left) where $P_{\rm P,Morph}<0.55$ (i.e. a reliable morphology could not be assigned). Left: MAGPI synthetic $g$, $r$, $i$-band colour images with white line (bottom right) showing a 5kpc physical scale and white circle (bottom left) showing the FWHM of the PSF. Right: Histogram showing the number of individual choices for the morphology of each galaxy. The mode of the input (ignoring IDKs) and posterior distributions are shown as vertical purple and red lines, respectively. Most galaxies without a reliable posterior mode morphology are either intermediate morphologies, poorly resolved or have complex (sub-)structures. In some cases, and depending on the individual classifier's performance, the model favours the relative abundances for the whole sample and assigns `E' (e.g. MAGPI1207227102).}\label{fig:snowflakes}
\end{figure}

In this section, we show example galaxies for which reliable classifications could not be obtained. Figure \ref{fig:snowflakes} shows example galaxies with $P_{\mu_{\rm Morph}}<0.55$, illustrating that most galaxies for which a reliable morphology classification could not be obtained are either intermediate morphologies, poorly resolved or have complex structures.

\begin{figure}
\includegraphics[width=9cm]{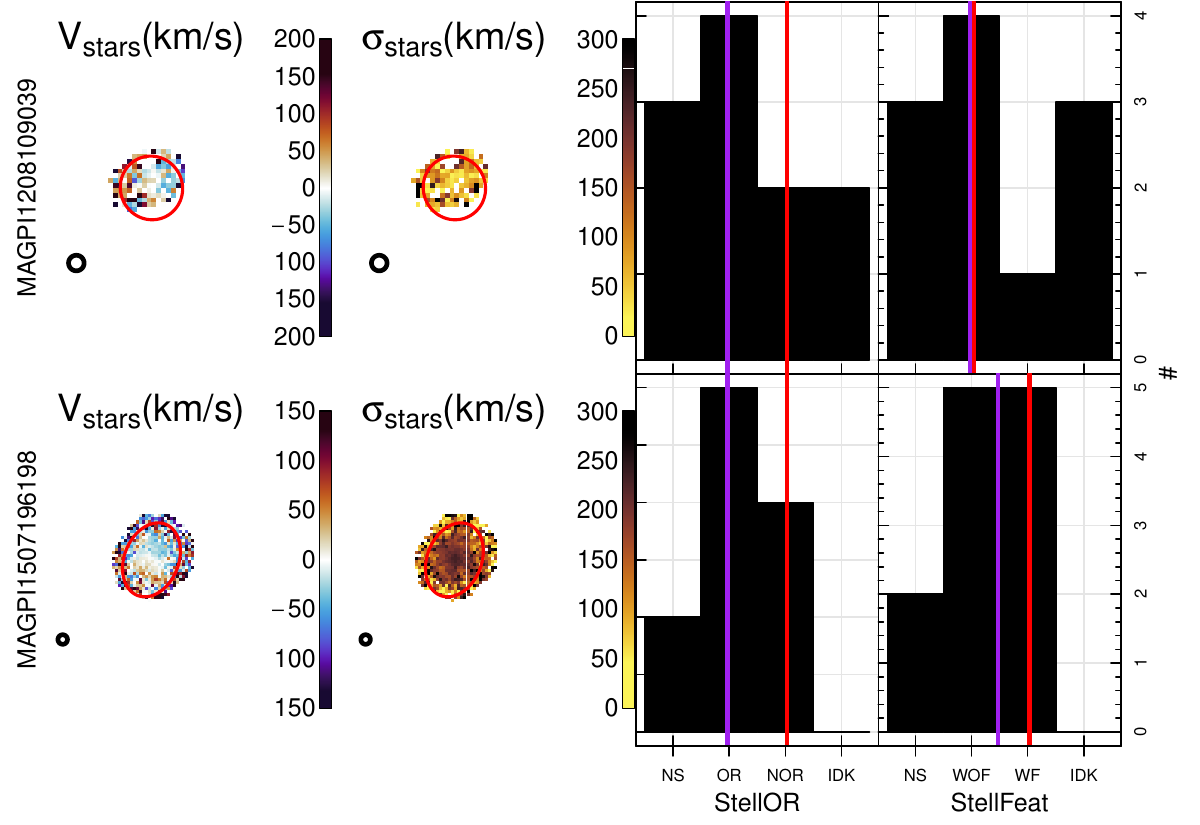}
\caption{MAGPI galaxies (as labelled on the left) where $P_{\rm P,StellOR}<0.55$ and $P_{\rm P,StellFeat}<0.55$ (i.e. reliable StellOR and StellFeat could not be assigned). Stellar velocity ($V_{\rm star}$, left) and dispersion ($\sigma$, left-centre) kinematic maps with black circle (bottom left on each panel) showing the PSF. Histogram showing the number of individual choices for StellOR (centre-left) and StellFeat (centre-right) for each galaxy. The mode of the input (ignoring IDKs) and posterior distributions are shown as vertical purple and red lines, respectively. Galaxies without reliable posterior modes for StellOR and StellFeat have complex stellar kinematic maps.}\label{fig:snowflakes_stell}
\end{figure}

Figure \ref{fig:snowflakes_stell} shows all galaxies for which $P_{\mu_{\rm StellOR}}<0.55$ and $P_{\mu_{\rm StellFeat}}<0.55$. In those cases, the stellar kinematic maps show complex stellar kinematic maps.

\begin{figure}
\includegraphics[width=9cm]{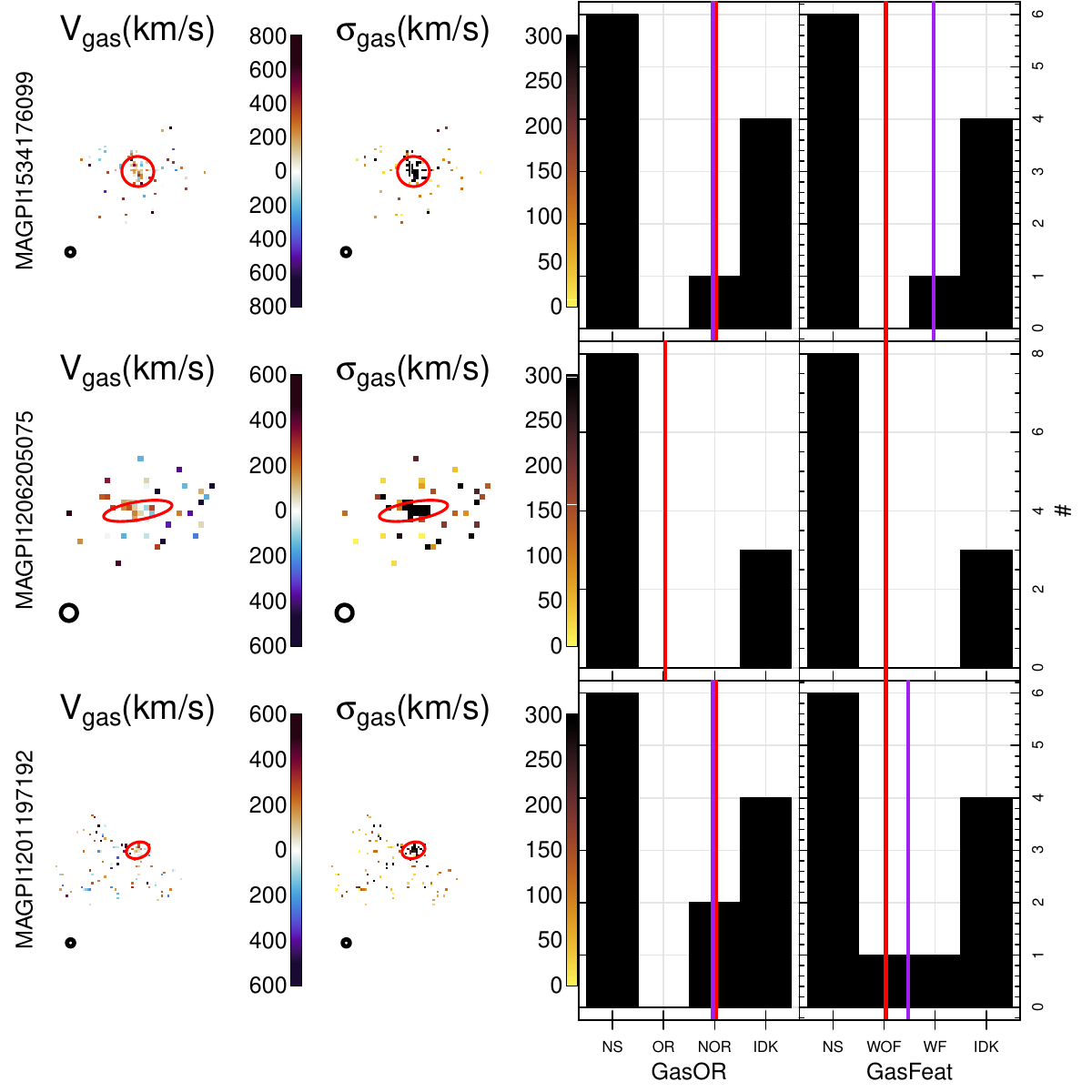}
\caption{Same as Figure \ref{fig:snowflakes_stell}, but for the ionised gas kinematics. Galaxies without reliable posterior modes for StellOR and StellFeat have sparse ionised gas kinematic maps.}\label{fig:snowflakes_gas}
\end{figure}

Figure \ref{fig:snowflakes_gas} shows all galaxies for which $P_{\mu_{\rm GasOR}}<0.55$ and $P_{\mu_{\rm GasFeat}}<0.55$. In those cases, the ionised gas kinematic maps are sparsely sampled.

\begin{figure}
\centering
\includegraphics[width=8.5cm]{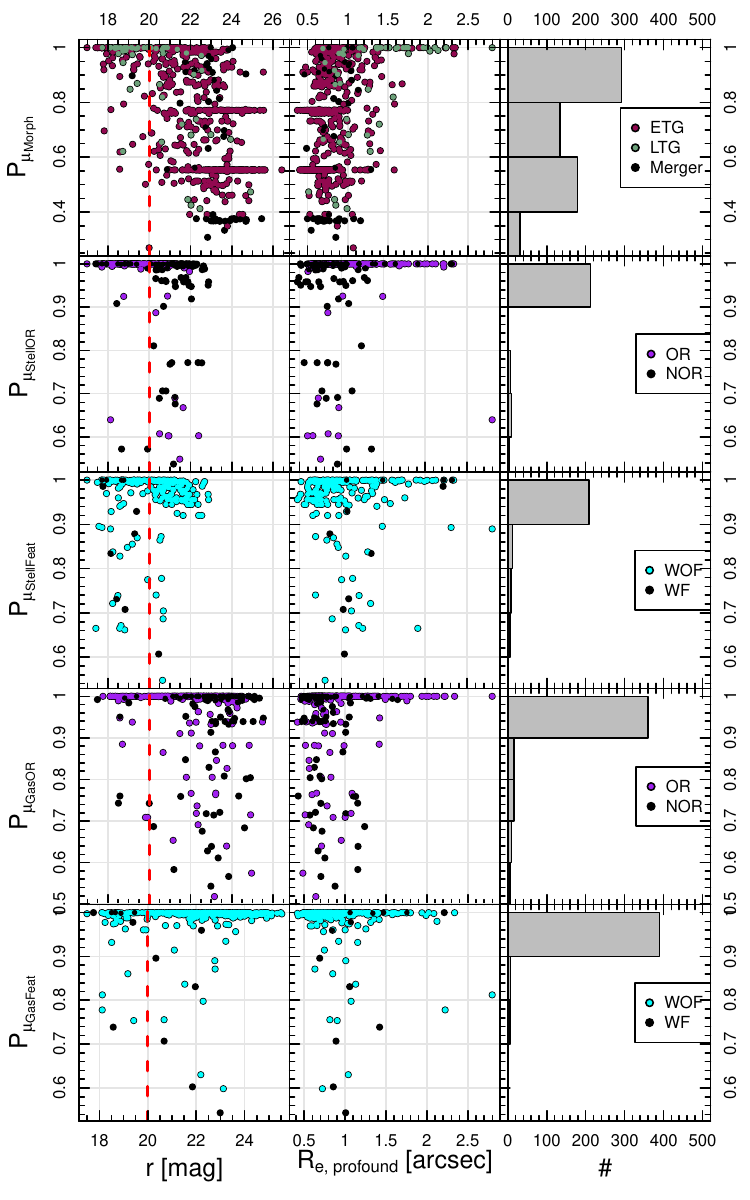}
\caption{Percentile function for the posterior mode probability $PF ({\mu})$ for Morph, StellOR, StellFeat, GasOR and GasFeat (top to bottom) on galaxy luminosity (i.e., $r$-band magnitude, left) and apparent effective radius ($R_{\rm e,profound}$, middle) as per {\sc ProFound}. Points are colour coded according to their posterior mode as per the legend in the right-most panel of each row. The distributions of $P_{\rm M}$ are shown as vertical histograms (right). Red dashed line shows the bright galaxies threshold $r\sim20$ mag. The scatter towards lower values of $P_{\mu}$ increases dramatically for objects fainter than $r\sim20$ mag and at small sizes ($R_{\rm e,profound} \sim 1-1.5$ depending on the feature).}\label{fig:rRe_fClass}
\end{figure}

In Figure \ref{fig:difficulties}, we show that the difficulty in classifying galaxies correlates with their brightness, size and the number of pixels shown in the relevant maps. Similarly, in Figure \ref{fig:rRe_fClass} we compare the posterior mode probabilities with the brightness and size of galaxies. In all but StellFeat and GasFeat and consistently with results shown in Figure \ref{fig:difficulties}, there is more scatter towards low posterior probabilities as the magnitude increases (i.e. faint galaxies have more low posterior probabilities classifications than bright ones). Similarly, there are proportionally more high confidence classifications for galaxies with larger apparent sizes ($R_e$ in arcsec) than for their smaller counterparts.

\section{Kinematic Classifications Summary}\label{sec:appendix_fields_kinclass}

This section shows a gallery of the MAGPI synthetic colour images for the MAGPI fields included in this work with visual and kinematic morphologies superimposed (Figure \ref{fig:kinclass_fields}). A small subset of these is shown in Figure \ref{fig:kinclass_fields_subs} in the main body of the manuscript.

\begin{figure*}
\centering
    \includegraphics[width=18cm]{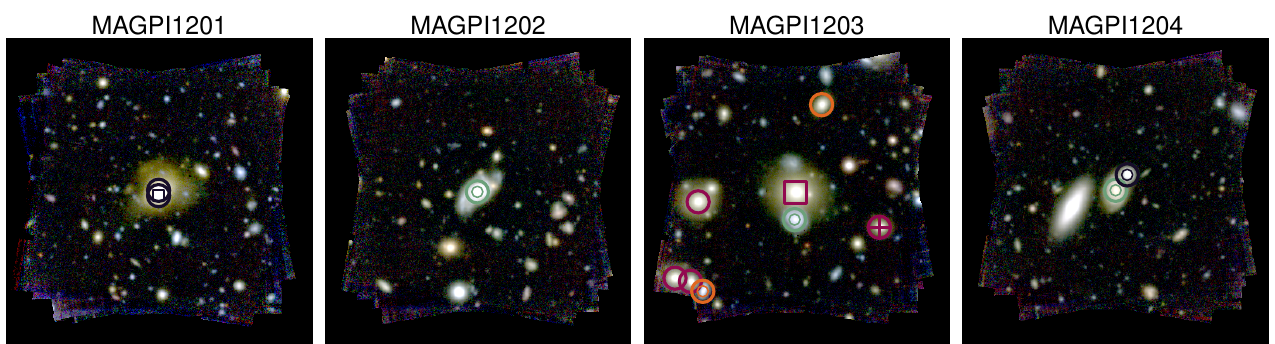}
    \includegraphics[width=18cm]{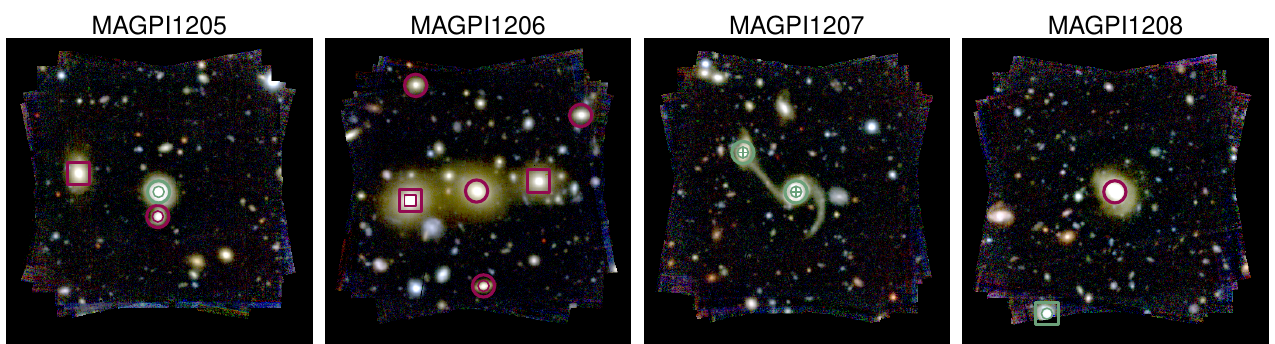}
    \includegraphics[width=18cm]{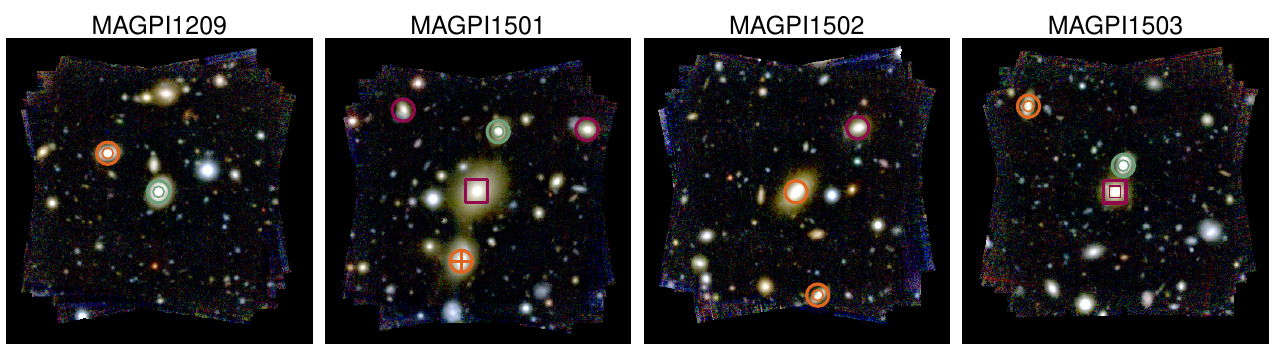}
    \includegraphics[width=18cm]{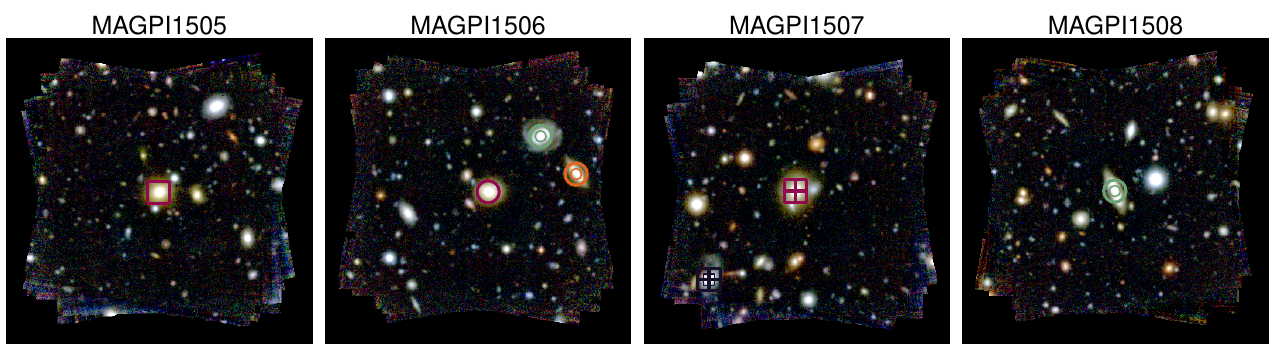}
\caption{Same as Figure \ref{fig:kinclass_fields_subs}, but for all MAGPI fields included in this work. For each field, a MUSE synthetic $g$, $r$, $i$ image is superimposed with gas and stellar visual kinematic morphologies: $\mu_{\rm StellOR}=$ OR (large thick circles), $\mu_{\rm GasOR}=$ OR (when available, small thin circles), $\mu_{\rm StellOR}=$ NOR (large thick squares), $\mu_{\rm GasOR}=$ OR (when available, small thin squares), $\mu_{\rm StellFeat}=$ WOF and $\mu_{\rm GasFeat}=$WOF (hollow symbols), $\mu_{\rm GasFeat}=$ WF and/or $\mu_{\rm StellFeat}=$ WF (thick and/or thin cross). Colours correspond to the mode of the posterior morphological classification $\mu_{\rm Morph}$ (see legend on the next page). In contrast to local surveys, there are many cases of NOR satellites and OR centrals, suggesting that while the kinematic diversity of galaxies is already established, the kinematic morphology-density relation is yet to be established.}\label{fig:kinclass_fields}
\end{figure*}
\begin{figure*}
\addtocounter{figure}{-1}
\centering
    \includegraphics[width=18cm]{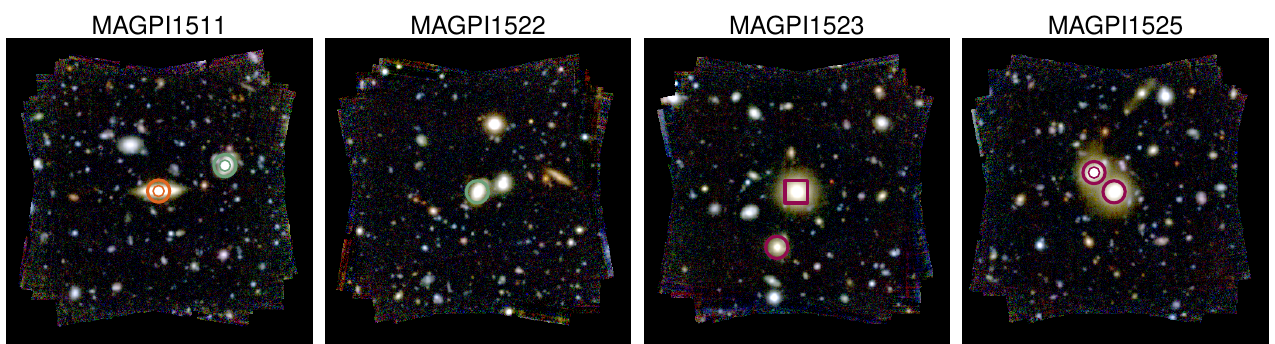}
    \includegraphics[width=18cm]{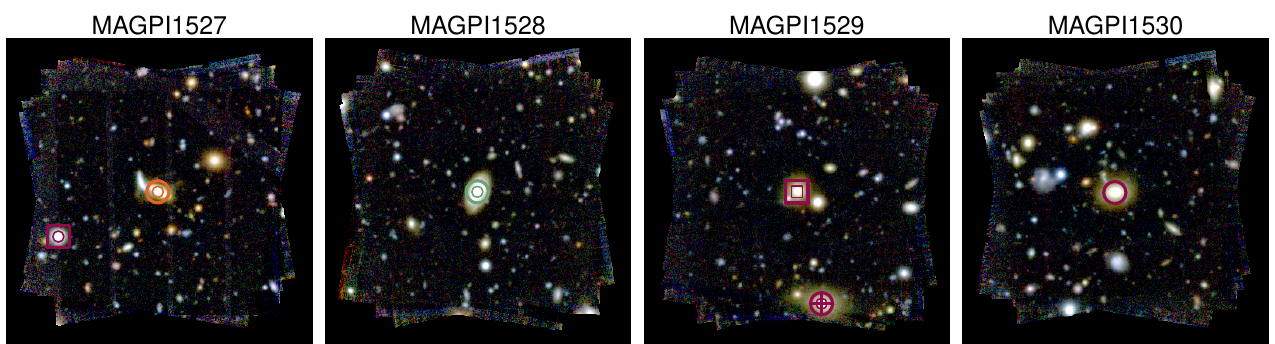}
    \includegraphics[width=18cm]{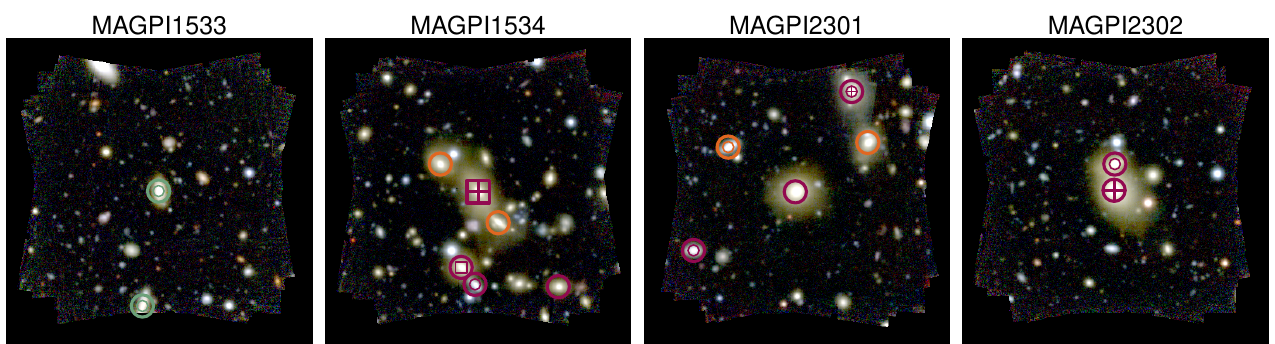}
    \includegraphics[width=18cm]{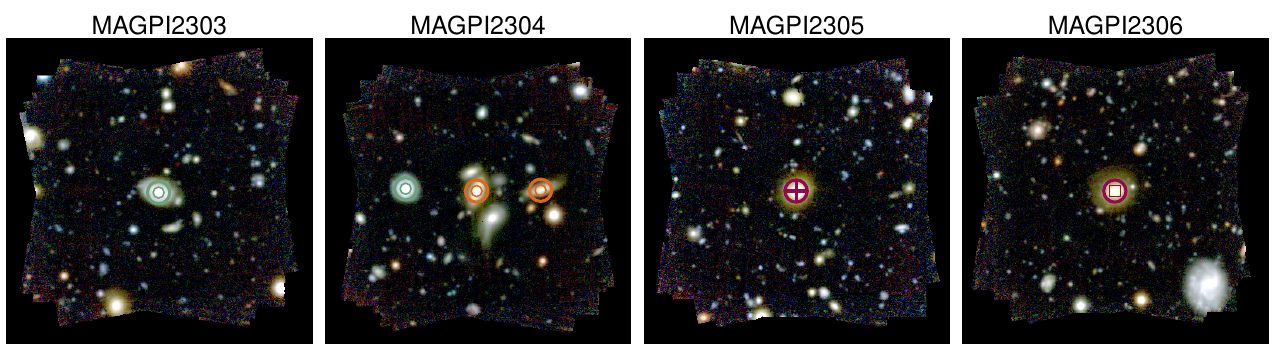}
    \includegraphics[width=18cm]{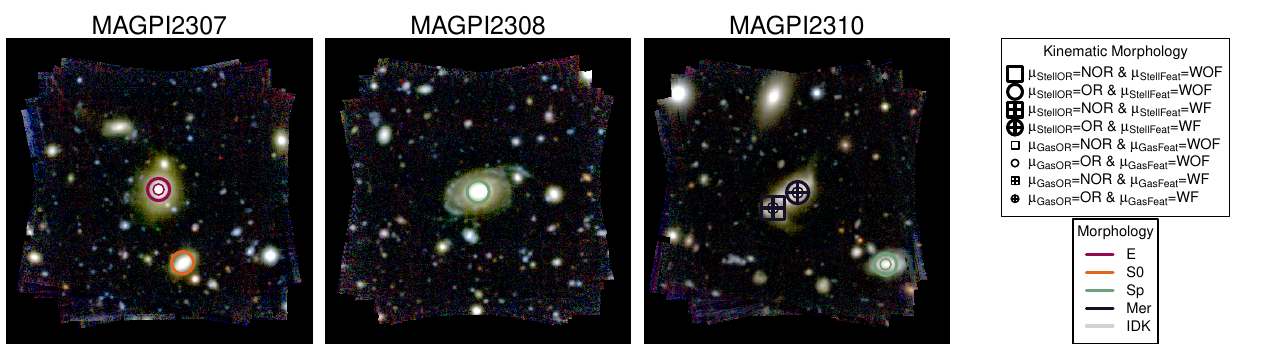}
\caption{Continued.}
\end{figure*}

\end{document}